\documentclass[12pt,eqsecnum]{article}
\usepackage[dvips]{graphicx}
\usepackage{amssymb}
\usepackage{amsmath}
\usepackage{epstopdf}
\usepackage{color}
\usepackage{soul}
\makeatletter

\@addtoreset{equation}{section}
\makeatletter

\newcommand{\qed}{\hbox{\rule[-2pt]{6pt}{6pt}}}
\newcommand{\D}{{\rm d}}

\newtheorem{lm}{Lemma}

\newcommand{\dalm}{\kern1pt\vbox{\hrule height 0.9pt\hbox{\vrule width
0.9pt\hskip 2.5pt\vbox{\vskip 5.5pt}\hskip 3pt\vrule width 0.3pt}\hrule height
0.3pt}\kern1pt}

\def\b2hat{ {\hat b}_2 }

\def\be {\begin{equation}}
\def\ee  {\end{equation}}
\def\bea {\begin{eqnarray}}
\def\eea {\end{eqnarray}}

\begin{document}

\begin{titlepage}
\vfill
\begin{flushright}
\today
\end{flushright}

\vfill
\begin{center}
\baselineskip=16pt
{\Large\bf 
Exact time-dependent states for throat quantized\\ toroidal AdS black holes \\
}
\vskip 0.5cm
{\large {\sl }}
\vskip 10.mm
{\bf Hideki Maeda${}^{a}$ and Gabor Kunstatter${}^{b}$} \\

\vskip 1cm
{
	${}^a$ Department of Electronics and Information Engineering, Hokkai-Gakuen University, Sapporo 062-8605, Japan\\
    ${}^b$ Department of Physics, University of Winnipeg and Winnipeg Institute for Theoretical Physics, Winnipeg, Manitoba, Canada R3B 2E9. \\
	\texttt{h-maeda@hgu.jp, g.kunstatter@uwinnipeg.ca}

     }
\vspace{6pt}
\today
\end{center}
\vskip 0.2in
\par
\begin{center}
{\bf Abstract}
\end{center}
\begin{quote}
We investigate exact non-stationary quantum states of vacuum toroidal black holes with a negative cosmological constant in arbitrary dimensions using the framework of throat quantization  pioneered by Louko and M\"akel\"a for Schwarzschild black holes. The system is equivalent to a harmonic oscillator on the half line, in which the central singularity is resolved quantum mechanically by imposing suitable boundary conditions that preserve unitarity. We identify two suitable families of exact time-dependent wave functions with Dirichlet or Neumann boundary conditions at the location of the classical singularity. We find that for highly non-stationary states of large-mass black holes, quantum fluctuations are not negligible in one family, while they are greatly suppressed in the other. The latter, therefore, may provide candidates for describing the dynamics of semi-classical black holes.\\
\vfill
\vskip 2.mm
\end{quote}
\end{titlepage}




\tableofcontents

\newpage

\section{Introduction}
The celebrated singularity theorems in general relativity~\cite{singularities} suggest that  the complete description of a black hole requires a quantum theory of gravity that can treat singularities in a consistent manner.
One would expect that singularities in classical general relativity are cured by quantum gravity in much the same way that the classical singularity is resolved in a simplified model of the quantized hydrogen atom.
This suggests a number of natural questions about properties of quantum black holes, such as their mass spectrum and interior structure, that are related to the end state of Hawking radiation. Apart from the important theoretical issues it raises, black-hole evaporation provides one of the few possible ultra-high-energetic astrophysical events that might one day be used to test a given quantum theory of gravity.
Unfortunately, we do not as yet have a complete quantum theory of gravity to provide us with precise answers to such questions.
Suggestive results may none the less be obtained using a midi-superspace approach in which one quantizes only a restricted class of highly symmetric spacetimes~\cite{midisuperspace}.

There are two different canonical methods for quantizing general relativity as a constrained dynamical system: Dirac quantization~\cite{Dirac} and reduced phase-space quantization.
In the former method, the constraints  in the Einstein field equations become operators acting on the wave function(al) of spacetime. The resulting Wheeler-DeWitt equation provides the basis for Dirac quantized gravity.
In reduced phase-space quantization, on the other hand, one first chooses a gauge, solves the classical constraint equations, and then puts the solutions to the constraints and gauge fixing conditions back into the action, which is then quantized on the resulting reduced phase space. One variation of this approach is to first do a suitable canonical transformation to a set of  phase space coordinates that separate out the gauge invariant modes and decouple them from the  modes that are ``pure gauge".  The gauge fixing procedure then becomes straightforward. This is the approach that we take in the following.

Although reduced phase-space quantization is a very natural way to quantize constrained dynamical systems,  it is technically hopeless to implement with full generality in Einstein's theory. 
It is nonetheless possible, {and highly instructive}, to carry out the reduced phase-space quantization of spherically symmetric vacuum spacetimes~\cite{kuchar94}. Classically, the Birkhoff theorem guarantees that the Schwarzschild solution is the unique one-parameter family of solutions in this system. 
Following Kucha\v{r}~\cite{kuchar94}, we take the Misner-Sharp quasi-local mass $M$~\cite{ms1964}, the areal radius ${R}$, and their conjugate momenta, $P_M$ and $P_R$, respectively, as canonical variables. The resulting constraint equations are trivial to solve: ${M}={\bf m}(t) $  is independent of the spatial coordinate, and $P_R= 0 $ on the constraint equations. In particular, they solve the constraint equations, resulting in a two-dimensional reduced phase space.
The resulting reduced action is simply
\begin{align}
I[{\bf m},{\bf p}]=\int \D t \biggl({\bf p}(t) \dot{{\bf m}}(t) - (N_+-N_-){\bf m}(t)\biggl), \label{intro}
\end{align}
where a dot denotes a derivative with respect to $t$~\cite{kuchar94}.
Here the dynamical variable is the spatially constant mode ${\bf m}(t)$ of the Misner-Sharp mass; i.e., its value on the constraint surface. Its conjugate momentum ${\bf p}$ is the Schwarzschild time separation between the two ends of the spatial slices at fixed time $t$, and the Hamiltonian is the mass ${\bf m}$ itself~\cite{kuchar94}.
The prescribed functions $N_\pm(t)$ are the values of the lapse at either end of the spatial slice and are not varied.
The Hamilton equation for ${\bf m}$ is simply ${\dot {\bf m}}=0$, which requires the spatially constant mode of  the Misner-Sharp mass, ${\bf m}$, to be constant in time as well.
This Kucha\v{r} reduction has been generalized in arbitrary dimensional spacetime with spherical, plane, or hyperbolic symmetry in the presence of a cosmological constant~\cite{dl2009}, and also in vacuum spherically symmetric Lovelock gravity~\cite{lovelock}, the most general metric theory of gravity yielding second-order field equations in arbitrary dimensions~\cite{kmt2013}.
The Kucha\v{r} action (\ref{intro}) in fact provides a firm {geometrical} foundation for studying  the quantum mechanics of the Schwarzschild-Tangherlini-type vacuum black holes in any theory admitting a Birkhoff's theorem.

Based in part on Kucha\v{r}' work, Louko and M\"akel\"a pioneered the method of throat quantization of the Schwarzschild black hole~\cite{LM96}.
They performed a canonical transformation from the Kucha\v{r} action (\ref{intro}) with $N_+=1$ and $N_-=0$ to a new action that describes the dynamics of the radius of the wormhole throat on a maximal slicing of the maximally extended Schwarzschild black-hole spacetime.
While $t$ is interpreted in the new action as the proper time on the wormhole throat, as pointed out by Louko and M\"akel\"a, one can choose the slicings so that $t$ is equal to the proper time in one of the asymptotically flat regions in the Schwarzschild spacetime.
Because the Hamiltonian for the throat dynamics is the Misner-Sharp mass, they quantized this system and identified its energy eigenvalues as mass eigenvalues of the Schwarzschild black hole~\cite{LM96}.

In a previous paper~\cite{km2014}, we generalized  results of Ref.~\cite{LM96} to arbitrary-dimensional vacuum black holes with spherical, planar, or hyperbolic symmetry, with or without a cosmological constant.
As in standard quantum mechanics, there are ambiguities in the choice of the operator ordering and boundary conditions.
In our study, we adopted Laplace-Beltrami operator ordering as a natural choice and imposed boundary conditions that preserve unitarity. As expected, we obtained discrete mass spectra for Schwarzschild-Tangherlini-type black holes in the stationary state~\cite{km2014}.
While the spectrum was obtained in the WKB approximation in most of the cases, we obtained an exact mass spectrum with a positive lower bound for asymptotically AdS toroidal black holes.
This exact spectrum was bounded below by a positive Planck scale number,  suggesting that the final state after the Hawking radiation of a toroidal black hole in a system with less symmetry would also be a Planck mass relic.

In quantum mechanics, it is natural to expect that an isolated system settles down to a stationary state in the far future due to dissipation of energy to the surroundings. 
Nonetheless, for suitably isolated systems, non-stationary exact quantum states also can have  physical relevance. The coherent states of the simple harmonic oscillator are a prime example. In quantum gravity, which is as yet not well understood at a fundamental level, the study of such exact dynamical states is particularly useful because it may shed insight into the expected semi-classical behaviour.

One key difference between the stationary and non-stationary states is that, while the expectation value of the mass is constant for both,  mass uncertainty exists only in the non-stationary case. This raises the following questions:
\begin{itemize}
\item How does the mass uncertainty change in time?
\item Does it have a maximum or minimum value?
\item How do we define the horizon of a quantum black hole in the dynamical case?
\item Once defined, how large are the fluctuations of the quantum horizon, and how far from the classical value is it located?
\item In the large-mass limit, do the quantum and classical horizons coincide, and do the quantum fluctuations become small?
\end{itemize}
We will address these questions in the framework of throat quantization of toroidal AdS black holes using two families of exact time-dependent wave functions.
One interesting result is that for highly non-stationary states of large-mass black holes, quantum fluctuations are not negligible in one family, while they are greatly suppressed in the other. Such states therefore provide candidates for describing the dynamics of semi-classical black holes.

The outline of the present paper is as follows.
In Sec.~\ref{sec:throat}, we review the throat quantization method and summarize our previous results for stationary states.
We also introduce the concepts of classical and quantum horizons as well as the semi-classical limit.
In Sec.~\ref{sec:wave} the wave functions for the toroidal black hole are presented.
In Secs.~\ref{sec:exact1} and \ref{sec:exact2}, we study the properties of quantum toroidal black holes described by two distinct families of wave functions.
Our conclusions and discussions are summarized in Sec.~\ref{sec:summary}.
Integral formulae involving Hermite polynomials that are used in the main text are given in Appendix A, while the details of computations are presented in Appendices B and C.
Our basic notation follows~\cite{wald}.
The convention for the Riemann curvature tensor is $[\nabla _\rho ,\nabla_\sigma]V^\mu ={R^\mu }_{\nu\rho\sigma}V^\nu$ and $R_{\mu \nu }={R^\rho }_{\mu \rho \nu }$.
The Minkowski metric is taken as diag$(-,+,\cdots,+)$, and Greek indices run over all spacetime indices.
We use a symbol $\kappa_n:=\sqrt{8\pi G_n}$, where $G_n$ is the $n$-dimensional Newton constant.
The Planck length and the Planck mass are defined by $\ell_{\rm p}:=(\hbar\kappa_n^2/c^3)^{1/(n-2)}$ and $m_{\rm p}:=(\hbar^{n-3}/\kappa_n^2c^{n-5})^{1/(n-2)}$, respectively.
We will use the Planck area and the Planck volume defined by $A_{\rm p}:=V_{n-2}^{(0)}\ell_{\rm p}^{n-2}$ and $V_{\rm p}:=V_{n-2}^{(0)}\ell_{\rm p}^{n-1}/(n-1)$, respectively, where $V_{n-2}^{(0)}$ represents volume of the $(n-2)$-dimensional subspace with toroidal symmetry.
Here we have kept the speed of light $c$ explicitly, but in the main text we set $c=1$.
In these units, we have $\hbar\kappa_n^2=\ell_{\rm p}^{n-2}$ and $\hbar=m_{\rm p}\ell_{\rm p}$.

\section{Throat quantization of vacuum symmetric black holes}
\label{sec:throat}

\subsection{Preliminaries}
We consider general relativity with a cosmological constant $\Lambda$ in arbitrary $n(\ge 3)$ dimensions. The action is given by 
\begin{align}
I=&\frac{1}{2\kappa_n^2}\int \D ^nx\sqrt{-g}(R-2\Lambda)+I_{\partial{\cal M}},\label{action}
\end{align}
where $I_{\partial{\cal M}}$ is the York-Gibbons-Hawking boundary term.
The above action gives the following vacuum Einstein field equations:
\begin{align} 
R_{\mu\nu}-\frac12 g_{\mu\nu}R+\Lambda g_{\mu\nu}=0. \label{beqL}
\end{align} 

We assume an $n$-dimensional warped product spacetime $({\cal M}^n,g_{\mu\nu}) \approx ({M}^2,g_{AB})\times ({K}^{n-2},\gamma_{ab})$, of which the most general metric is given by 
\begin{align}
g_{\mu\nu}(x)\D x^\mu \D x^\nu=g_{AB}({\bar y})\D {\bar y}^A \D {\bar y}^B+r({\bar y})^2\gamma_{ab}(z)\D z^a\D z^b,
\label{eq:structure}
\end{align}
where indices run as $A,B=0,1$ and $a,b=2,3,\cdots,n-1$.
Here $({M}^2,g_{AB})$ is the most general two-dimensional Lorentzian manifold, and $({K}^{n-2},\gamma_{ab})$ is the $(n-2)$-dimensional maximally symmetric space with its curvature $k=1, 0, -1$; namely, the Riemann tensor ${}^{(n-2)}R^{ab}_{~~cd}$ on $({K}^{n-2},\gamma_{ab})$ is given by
\begin{eqnarray}
{}^{(n-2)}R^{ab}_{~~cd}=k(\delta^a_c \delta^b_d-\delta^a_d \delta^b_c).
\end{eqnarray}
Note that, in three dimensions, $({K}^{n-2},\gamma_{ab})$ is one-dimensional and $k=0$ necessarily holds.
We assume that $({K}^{n-2},\gamma_{ab})$ is compact.
$r({\bar y})$ is a scalar on $({M}^2,g_{AB})$, called the areal radius because $r^{n-2}$ is proportional to the area of a symmetric $(n-2)$-surface generated by the spatial Killing vectors on $({K}^{n-2},\gamma_{ab})$.

The generalized Misner-Sharp quasi-local mass~\cite{ms1964,mn2008} in this system is defined by
\begin{align}
M:=& \frac{(n-2)V_{n-2}^{(k)}}{2\kappa_n^2}r^{n-3}\biggl(\frac{r^2}{l^2}+(k-(Dr)^2)\biggl),\label{qlm}
\end{align}  
where $D_A$ is the covariant derivative on $({M}^2,g_{AB})$, $(Dr)^2:=(D_A r)(D^A r)$, and the AdS radius $l$ is defined by 
\begin{align}
l^2:=&-\frac{(n-1)(n-2)}{2\Lambda}. \label{lambdatil}
\end{align}  
The constant $V_{n-2}^{(k)}$ represents the volume of $({K}^{n-2},\gamma_{ab})$.
$M$ reduces to the ADM mass at spacelike infinity in the asymptotically flat spacetime~\cite{mn2008,hayward1996}.

\subsection{Throat dynamics of vacuum black holes}
\label{sec:ClassicalThroatDynamics}
In the vacuum case, the Einstein equations show that the mass function $M$ is constant.
Then, in the case of $k\Lambda \le 0$, the general vacuum solution is the following generalized Schwarzschild-Tangherlini solution:
\begin{align}
\D s^2=&-f(r)\D t^2+f(r)^{-1}\D r^2+r^2\gamma_{ab}\D z^a \D z^b, \label{f-vacuum} \\
f(r) =&k-\frac{2\kappa_n^2M}{(n-2)V_{n-2}^{(k)}r^{n-3}}+\frac{r^2}{l^2}.
\end{align}
For $k\Lambda > 0$, by contrast, the general vacuum solution consists of the solution (\ref{f-vacuum}) and the Nariai ($k=1$) or anti-Nariai ($k=-1$) direct product solution in which the areal radius $r$ is constant.
\begin{figure}[htbp]
\begin{center}
\includegraphics[width=0.5\linewidth]{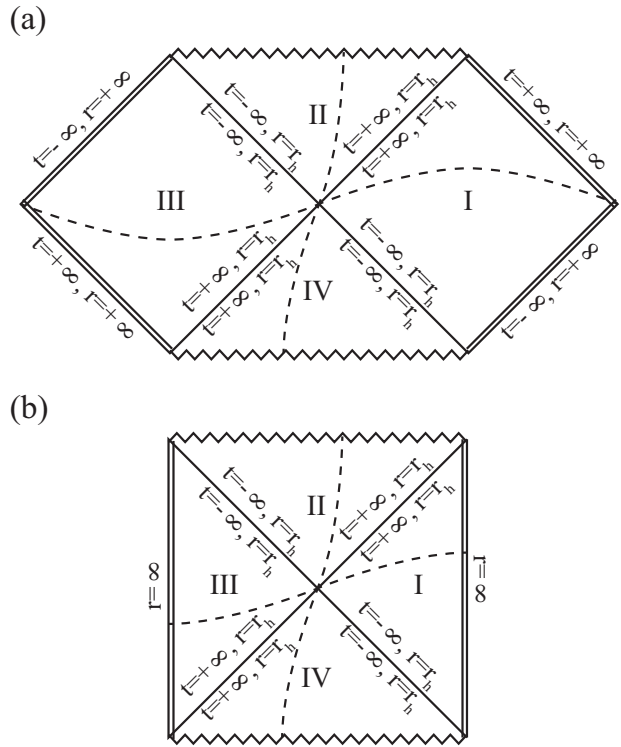}
\caption{\label{SingleBH} Penrose diagrams for the Schwarzschild-Tangherlini-type black hole (\ref{f-vacuum}) with a single horizon with (a) $\Lambda=0$ and (b) $\Lambda<0$. While a zigzag line represents a curvature singularity at $r=0$, a dashed curve in each portion of the spacetime represents a constant $t$ hypersurface.
In (a), a double line represents null infinity. In (b), it represents both null and spacelike infinities.}
\end{center}
\end{figure}

In the generalized Schwarzschild-Tangherlini spacetime (\ref{f-vacuum}), there is a central curvature singularity at $r=0$ and a Killing horizon located at $r=r_{\rm h}$ such that $f(r_{\rm h})=0$.
Hence, the mass-horizon relation is given by
\begin{align}
M=\frac{(n-2)V_{n-2}^{(k)}r_{\rm h}^{n-3}}{2\kappa_n^2}\biggl(k+\frac{r_{\rm h}^2}{l^2}\biggl). \label{mass-horizon}
\end{align}
The Penrose diagrams for the generalized Schwarzschild-Tangherlini black hole (\ref{f-vacuum}) with a single horizon are drawn in Fig.~\ref{SingleBH}.
(The black hole does not possess multiple horizons in the case of $k=1$ or $k=0$.)

The maximally extended black-hole spacetime drawn in Fig.~\ref{SingleBH} has the structure of a wormhole whose throat is located inside the event horizon $r=r_{\rm h}$.
The areal radius of this wormhole throat evolves with time. We  represent it by $r=a(\tau)$, where $\tau$ is the proper time on the throat. Its time evolution is generated by the following Hamiltonian $H[a(\tau),p_a(\tau)]$:
\begin{align}
H=&\frac{(n-2)V_{n-2}^{(k)}a^{n-3}}{2\kappa_n^2}\biggl\{\biggl(\frac{(n-2)V_{n-2}^{(k)}}{\kappa_n^2}\biggl)^{-2}p_a^2a^{-2(n-3)}+k+\frac{a^2}{l^2}\biggl\}, \label{hamiltonian}
\end{align}
where $p_a$ is the momentum conjugate of $a$.
The behaviour of the throat radius as described by the resulting solutions $a(\tau)$ is as follows: the throat radius starts to increase from the white-hole singularity at $a=0$ until it reaches its turning point at  the bifurcation $(n-2)$-surface $a=r_{\rm h}$.  The throat radius then decreases from  its maximum value $a=r_{\rm h}$, and finally reaches the black-hole singularity at $a=0$.
The maximum value $r_{\rm h}$ of $a$ coincides with the horizon radius for any given value of $M$. 
Since the horizon radius as given in Eq.~(\ref{mass-horizon}) increases indefinitely as $M$ increases, the domain of $a$ is $0\le a <\infty$, while the domain of $p_a$ is $-\infty<p_a<\infty$.

Incidentally, this provides a fundamental motivation for quantization using throat variables, rather than the Misner-Sharp mass, as done by Kucha\v{r}. As described above, the physics prescribes the range of the phase-space variables $a$ and $p_a$, which in turn determine the spectrum of the energy. If one tries to quantize the energy directly, its conjugate variable is effectively time, which presumably is a continuous variable in $(-\infty,+\infty)$, so that one would naturally obtain a continuous spectrum of plane wave solutions with mass unbounded below. In this context, a continuous spectrum $(-\infty,+\infty)$ for the mass $M$ is natural also, because the Kucha\v{r} reduced action (\ref{intro}) does not retain any information about the fact that the spacetime is that of a black hole. Moreover, if one tries to restrict the range of $M$ to ${\mathbb R}^+$, then as with a free particle on the half line, its conjugate would not exist as a self-adjoint operator.

A fundamental feature of throat quantization is that, as shown for Schwarzschild black holes in Ref.~\cite{LM96}, the  corresponding Hamiltonian (\ref{hamiltonian}) can be obtained by a canonical transformation from the Kucha\v{r} action (\ref{intro}) with $N_+=1$ and $N_-=0$, in which the Hamiltonian is the mass $M$ of the black hole and the time $t$ coincides with the proper time $\tau$ of an observer at rest on the throat of the wormhole.
One then quantizes the throat dynamics, whose action is given by 
\begin{align}
I=\int \D t\biggl(p_a{\dot a}-H[a,p_a]\biggl). \label{newaction}
\end{align}
We emphasize that $t$ in the new action (\ref{newaction}) is originally the Kucha\v{r}' asymptotic time variable, which is the proper time of an observer at spacelike infinity in the spacetime (\ref{f-vacuum}), so the expectation values of physical quantities are those measured by such an observer.
The existence of such a canonical transformation provides the required connection between the throat variables and  the geometrodynamics of the full diffeomorphism invariant spacetime that underlies the Kucha\v{r} action.

\subsection{Hamiltonian operator}
We now proceed to quantize (\ref{newaction}) using Schr\"odinger quantization.
Replacing $p_a$ in the Hamiltonian (\ref{hamiltonian}) by ${\hat p}_a=-i\hbar \partial/\partial a$, we obtain the Schr\"odinger equation for the wave function $\Psi(t,a)$:
\begin{align}
{\hat H}\Psi =i\hbar\frac{\partial\Psi}{\partial t}. \label{Schro}
\end{align}  
Adopting Laplace-Beltrami operator ordering~\cite{zanelli1986}, we obtain the following Hamiltonian operator~\cite{km2014}:
\begin{align}
{\hat H} =&\frac{(n-2)V_{n-2}^{(k)}}{2\kappa_n^2}\biggl\{\frac{a^{n-1}}{l^2}+ka^{n-3}-\biggl(\frac{\kappa_n^2}{(n-2)V_{n-2}^{(k)}}\biggl)^2\frac{\hbar^2}{a^{(n-3)/2}}\frac{\partial}{\partial a}\biggl(\frac{1}{a^{(n-3)/2}}\frac{\partial }{\partial a}\biggl)\biggl\}.
\label{H_LB}
\end{align}  
This operator is Hermitian symmetric with respect to the inner product defined by
\begin{align}
\langle \Psi|\Phi\rangle:=\int^\infty_0\Psi^*\Phi\mu(a)\D a.
\end{align}  
for two arbitrary wave functions $\Psi$ and $\Phi$, with measure  $\mu(a)=a^{(n-3)/2}$. Louko and M\"akel\"a considered the more general measure $\mu(a) = a^\beta$~\cite{LM96}, {and corresponding Hermitian ordering of the momentum term in the Hamiltonian}.

Defining $x:=a^{(n-1)/2}$, we rewrite ${\hat H}$ as
\begin{align}
{\hat H} =&\frac{(n-2)V_{n-2}^{(k)}}{2\kappa_n^2}\biggl\{\frac{x^2}{l^2}+kx^{2(n-3)/(n-1)} -\hbar^2\biggl(\frac{(n-1)\kappa_n^2}{2(n-2)V_{n-2}^{(k)}}\biggl)^2\frac{\partial ^2}{\partial x^2}\biggl\}. \label{H-x}
\end{align}  
Note that $x^2 = a^{n-1}$ is proportional to the Euclidean (or special relativistic) volume of a throat with radius $a$.
Comparing Eq.~(\ref{Schro}) with 
\begin{align}
\biggl(-\frac{\hbar^2}{2m}\frac{\partial ^2}{\partial x^2}+V(x)\biggl)\psi=i\hbar\frac{\partial\Psi}{\partial t}, \label{Schro-eq}
\end{align}  
we identify the effective mass and potential as
\begin{align}
V(x)\equiv &\frac{(n-2)V_{n-2}^{(k)}}{2\kappa_n^2}\biggl(\frac{x^2}{l^2}+kx^{2(n-3)/(n-1)}\biggl),\label{p-potential}\\
m\equiv &\frac{4(n-2)V_{n-2}^{(k)}}{(n-1)^2\kappa_n^2}. \label{p-mass}
\end{align} 
In terms of $x$, the inner product becomes simple:
\begin{align}
\langle \Psi|\Phi\rangle:=\int^\infty_0\Psi^*\Phi\D{x}.
\end{align}  
It is important to note that since the exponent in the second term of (\ref{p-potential}) is less than 2 for all values of $n$, the potential $V(x)$ is bounded below for $k=-1$ with $\Lambda<0$.

\subsection{Self-adjointness of the Hamiltonian operator}
We require the quantum system  to obey unitarity:
\begin{align}
\frac{\D}{\D t}\langle \Psi|\Phi\rangle=0\,.
\end{align}  
This requires the Hamiltonian operator ${\hat H}$ to be self-adjoint on the domain of $0 \le x<\infty$ or to have a self-adjoint extension, which guarantees reality of the eigenvalues. 
In our previous paper~\cite{km2014}, we showed that the Hamiltonian operator (\ref{H-x}) on the half line $x\in [0,\infty)$ admits an infinite number of self-adjoint extensions. We now summarize the arguments that lead to this conclusion.
(See Ref.~\cite{robin bcs} for general discussions of self-adjoint extensions of operators.)

Starting with Eqs.~(\ref{Schro}) and (\ref{H-x}) and integrating by parts with fall-off condition $\Psi\to 0$ as $x\to \infty$, one obtains
\begin{align}
\frac{\D}{\D t}\langle \Psi|\Phi\rangle=\frac{i\hbar}{2m} \biggl(\frac{\partial\Psi^*}{\partial x}\Phi-\Psi^*\frac{\partial\Phi}{\partial x}\biggl)\biggl|_{x=0}.
\end{align}  
The right-hand side of the above equation must vanish for all states in the Hilbert space in order to ensure unitarity of the system.
Thus, the wave functions must obey the following boundary condition:
\begin{align}
\Psi(t,0)+L\frac{\partial\Psi}{\partial x}(t,0)=0,
\label{eq:robin bc}
\end{align}  
where $L$ is a real constant. 
$L=0$ and $L=\infty$ correspond to Dirichlet and Neumann boundary conditions, respectively.
The remaining values of $L$ correspond to Robin boundary conditions~\cite{robin bcs}. 
Since one obtains inequivalent but well-defined quantum theories for different real values of $L$, the Hamiltonian operator (\ref{H-x}) on the half line $x\in [0,\infty)$ admits an infinite number of self-adjoint extensions.

We note that in the case of quantization on the whole line $x\in (-\infty,\infty)$, Ehrenfest's theorem holds;
\begin{align}
m\frac{\D^2\langle x\rangle}{\D t^2}=-\biggl\langle \frac{\D V}{\D x}\biggl\rangle,
\end{align}
so that $\langle x\rangle$ follows quantum corrected classical orbits (since $\langle \frac{\D V}{\D x}\rangle\neq \left.\frac{\D V}{\D x}\right|_{x=\langle x\rangle}$). 
The above equation is modified in the case of the half-line $x\in [0,\infty)$.
Under the assumption that the surface terms at infinity vanish, and again using the Schr\"odinger equation~(\ref{Schro-eq}) and integration by parts, we obtain
\begin{align}
m\frac{\D^2\langle x\rangle}{\D t^2}=&-\frac{\hbar^2}{4m} \biggl(\frac{\partial^2\Psi^*}{\partial x^2}\Psi+\Psi^*\frac{\partial^2\Psi}{\partial x^2}-2\frac{\partial\Psi^*}{\partial x}\frac{\partial\Psi}{\partial x}\biggl)\biggl|_{x=0} -\biggl\langle \frac{\D V}{\D x}\biggl\rangle.
\end{align}
Therefore, in general, the expectation value of $x$ does not follow classical orbits due to possible boundary effects\footnote{This is related to the fact that the conjugate operator $\hat{p}=-i\hbar \partial/\partial x$ does not exist on the half line as a self-adjoint operator, nor does it have self-adjoint extensions.}.
(See Ref.~\cite{maeda2015} for the case of $\langle x^q\rangle$, where $q$ is a positive constant.)
{We note, however, that for $q=2$, one has
\begin{align}
m\frac{\D^2\langle x^2\rangle}{\D t^2}=& \frac{\hbar^2}{2m}
    \biggl\{x\biggl( \frac{\partial^2 \Psi^*}{\partial x^2}\Psi + \Psi^* \frac{\partial^2 \Psi}{\partial x^2} \biggl)
-2\biggl(\frac{\partial \Psi^*}{\partial x}\Psi + \Psi^*\frac{\partial \Psi}{\partial x}\biggl)\biggl\}\biggl|_{x=0}\nonumber\\
  & + \int^\infty_0 \D x\biggl(\frac{2\hbar^2}{m} \frac{\partial \Psi^*}{\partial x}\frac{\partial \Psi}{\partial x}
 -2 x \Psi^* \frac{\D V}{\D x}\Psi\biggl). 
\end{align}
Assuming that the wave function and its first and second derivatives are finite at $x=0$, the boundary terms vanish for Dirichlet and Neumann boundary conditions. These are the two boundary conditions that we consider  in what follows. }

\subsection{{Stationary states of the quantum black hole}}
While the classical relation between the mass parameter $M$ and the horizon radius $r_{\rm h}$ is given by Eq.~(\ref{mass-horizon}), {the mass spectrum of the quantum black hole is discrete in throat quantization}.
In a previous paper~\cite{km2014}, we studied the mass spectrum for asymptotically flat and AdS black holes in stationary states.

An intriguing result in Ref.~\cite{km2014} is that the presence of a negative cosmological constant $\Lambda$ drastically changes the spectrum: While the mass of the black hole is equally spaced in the asymptotically AdS case, entropy is equally spaced in the asymptotically flat case.
In this analysis, the entropy spectrum is obtained from the classical relation between the mass $M$ and entropy $S$ of a black hole, which is 
\begin{align}
S=\frac{2\pi}{\kappa_n^2}V_{n-2}^{(1)}\biggl(\frac{2\kappa_n^2\langle M\rangle_N}{(n-2)V_{n-2}^{(1)}}\biggl)^{(n-2)/(n-3)} \label{entropy-k=1}
\end{align}
for $k=1$ and $\Lambda=0$, and 
\begin{align}
S=&\frac{2\pi}{\kappa_n^2}V_{n-2}^{(0)}\biggl(\frac{2\kappa_n^2l^2\langle M\rangle_N}{(n-2)V_{n-2}^{(0)}}\biggl)^{(n-2)/(n-1)} \label{entropy-k=0}
\end{align}
for $k=0$ and $\Lambda<0$.
The relation (\ref{entropy-k=0}) also holds for black holes with $k=\pm 1$ and $\Lambda<0$ in the large-mass limit.
In the following subsections, we will summarize the results in Ref.~\cite{km2014}.

\subsubsection{Asymptotically flat black hole}
For asymptotically flat black holes ($k=1$) with $\Lambda=0$, the WKB approximation yields a mass spectrum that is not equally spaced. Instead, it is the entropy that is equally spaced~\cite{km2014}:
\begin{align}
 S\simeq  &4\pi^{3/2}\hbar \frac{\Gamma(\frac{n-1}{2(n-3)}+\frac12)}{\Gamma(\frac{n-1}{2(n-3)})}N=:S_N,\label{spectrum-flat}
\end{align}
where $N$ is a large integer.
This is independent of the extension parameter $L$ in the boundary condition (\ref{eq:robin bc}).
Some particular cases of the above are 
\begin{eqnarray}
S_N=
\left\{ \begin{array}{ll}
4\pi^{3/2}\hbar N\frac{\Gamma(2)}{\Gamma(3/2)}=4\pi \hbar N~~(n=4)\\
4\pi^{3/2}\hbar N\frac{\Gamma(3/2)}{\Gamma(1)} =\pi^2 \hbar  N~(n=5)\\
4\pi^{3/2}\hbar \frac{\Gamma(1)}{\Gamma(1/2)}N=2\pi\hbar N~~(n\to \infty).
\end{array} \right.
\end{eqnarray}
The relation (\ref{entropy-k=1}) with the result (\ref{spectrum-flat}) means that for large $N$ the mass spectrum behaves as
\begin{align}
\langle M\rangle_N\propto N^{(n-3)/(n-2)}. \label{spectrum-mass-flat}
\end{align}

\subsubsection{Asymptotically AdS black hole}
In the presence of negative $\Lambda$, the mass spectrum for toroidal black holes ($k=0$) under the Dirichlet or Neumann boundary condition can be obtained exactly as
\begin{align}
\langle M\rangle_N=\frac{(n-1)\hbar}{2l}\biggl(N+\frac12\biggl) \quad (N=0,1,2,\cdots), \label{EN:toroidal}
\end{align}  
where even (odd) $N$ corresponds to the Neumann (Dirichlet) boundary condition~\footnote{The integer $N$ in Eq.~(3.12) in Ref.~\cite{km2014} with Dirichlet boundary conditions corresponds to half of the integer $N$ in Eqs.~(\ref{EN:toroidal}) and (\ref{spectrum-AdS}) in the present paper.}.
On the other hand, the spectrum for black holes with $k=\pm 1$ has only been obtained in the WKB approximation:
\begin{align}
\langle M\rangle_N \simeq \frac{(n-1)\hbar}{2l}N, \label{spectrum-AdS}
\end{align}  
where $N$ is a large integer.
The spectrum (\ref{spectrum-AdS}) for large $N$ is independent of the extension parameter $L$ in the boundary condition (\ref{eq:robin bc}).
\begin{figure}[htbp]
\begin{center}
\includegraphics[width=0.65\linewidth]{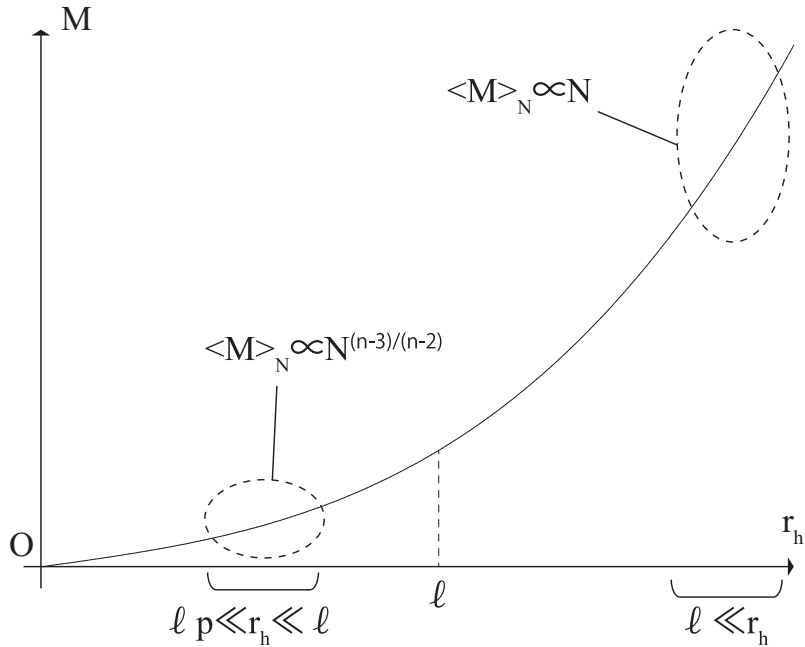}
\caption{\label{fig-AdSspectrum} Mass-horizon relation (\ref{mass-horizon}) for spherical AdS black holes ($k=1$). The mass spectrum is given by Eq.~(\ref{spectrum-flat}) in the region of $\ell_{\rm p}\ll r_{\rm h}\ll l$ and by Eq.~(\ref{spectrum-AdS}) in the region of $r_{\rm h}\gg l$.}
\end{center}
\end{figure}

At first glance, the two results (\ref{spectrum-mass-flat}) and (\ref{spectrum-AdS}) appear to be inconsistent in the limit of vanishing cosmological constant $\Lambda\to0$ ($l\to\infty$) for spherical AdS black holes ($k= 1$).
However, this apparent discrepancy can easily be understood by noting the presence of two different length scales, namely the Planck length $\ell_{\rm p}$ and the AdS radius $l$.
The mass-horizon relation (\ref{mass-horizon}) shows that a spherical AdS black hole is approximated by an asymptotically flat black hole if the horizon radius $r_{\rm h}$ is much smaller than the AdS radius $l$, while it is approximated by the toroidal AdS black hole ($k=0$) if $r_{\rm h}$ is much larger than $l$.
Thus, the spectrum of the spherical AdS black hole is given by Eq.~(\ref{spectrum-flat}) in the region of $\ell_{\rm p}\ll r_{\rm h}\ll l$ and by Eq.~(\ref{spectrum-AdS}) in the region of $r_{\rm h}\gg l$. (See Fig.~\ref{fig-AdSspectrum}.)
In the limit of zero cosmological constant ($l\to\infty$), the black-hole spacetime becomes asymptotically flat so that the region $r_{\rm h}\gg l$ disappears in Fig.~\ref{fig-AdSspectrum} and the mass spectrum (\ref{spectrum-flat}) is valid in the region of $r_{\rm h}\gg \ell_{\rm p}$\footnote{The authors thank Shunichiro Kinoshita for this interpretation of our result.}.

\subsection{Horizons of the quantum toroidal black hole}

Now let us focus on the toroidal black hole ($k=0$).
Classically, the horizon radius $r=r_{\rm h}$ with mass $M$ is given by
\begin{align}
r_{\rm h}=&\biggl(\frac{2\kappa_n^2l^2M}{(n-2)V_{n-2}^{(0)}}\biggl)^{1/(n-1)} =l_{\rm p}\biggl(\frac{2l^2M}{(n-2)V_{n-2}^{(0)}l_{\rm p}^2m_{\rm p}}\biggl)^{1/(n-1)}. \label{c-horizon}
\end{align}
How, then, does one define the location of the ``horizon'' for a quantum black hole?

\subsubsection{``Classical" horizon}
One option is simply to use Eq.~(\ref{c-horizon}) with $M$ replaced by the mass expectation value, namely
\begin{align}
r_{\rm C}:=l_{\rm p}\biggl(\frac{2l^2\langle M\rangle_N}{(n-2)V_{n-2}^{(0)}l_{\rm p}^2m_{\rm p}}\biggl)^{1/(n-1)}. \label{q-horizon1}
\end{align}
In the stationary state, there is no mass fluctuation, so that $r_{\rm C}$ is fixed without ambiguity.
For this reason, we call $r_{\rm C}$ the areal radius of the ``classical'' horizon.
The surface area of the classical horizon is then defined by 
\begin{align}
A_{\rm C}:=V_{n-2}^{(0)}r_{\rm C}^{n-2}=A_{\rm p}\biggl(\frac{2l^2\langle M\rangle_N}{(n-2)V_{n-2}^{(0)}l_{\rm p}^2m_{\rm p}}\biggl)^{(n-2)/(n-1)}.\label{def-Ac}
\end{align}

However, in non-stationary states, mass fluctuations $\delta \langle M\rangle_N$ cause the fluctuations of the radius $\delta r_{\rm C}$ and the surface area $\delta A_{\rm C}$ of the classical horizon.

\subsubsection{Quantum horizon}
Another option to define the location of the horizon for a dynamical quantum black hole comes from the throat dynamics. 
As explained in Sec.~\ref{sec:ClassicalThroatDynamics},  the classical throat radius $a(t)$ starts to grow from zero corresponding to the white-hole singularity, turns to decrease after reaching the maximum value at the bifurcation $(n-2)$-surface, and finally becomes zero again at the black-hole singularity. (See Fig.~\ref{SingleBH}(b).)
Because the maximum value of $a(t)$ at the bifurcation $(n-2)$-surface corresponds to the horizon radius classically, one may identify the location of the horizon of a quantum black hole by the maximum value of $\langle a\rangle_N(t)$, namely
\begin{align}
r_{\rm Q}:=\max_{t\in\mathbb{R}}\langle a\rangle_N(t).
\end{align}
We call $r_{\rm Q}$ the areal radius of the ``quantum'' horizon. 
Note that such an option is not available for stationary states, since of course the expectation values do not change with time. 
One can instead  identify the quantum turning point in the semi-classical limit with the value of $a$ (far from the origin  in the case of the half line) where the probability density is greatest, since for the classical system the probability of finding the particle is greatest near the point where the velocity vanishes.

However, using the above arguments, one may also choose to define the size of the black hole by the maximum value, not of the areal radius $a(t)$, but instead of the surface area $A(t):=V_{n-2}^{(0)}a^{n-2}$.
Then, the surface area of the quantum black hole is defined by 
\begin{align}
A_{\rm Q}:=\max_{t\in\mathbb{R}}\langle A\rangle_N(t),
\end{align}
where $A_{\rm Q}\ne V_{n-2}^{(0)}r_{\rm Q}^{n-2}$ holds in general.
The fluctuations of $r_{\rm Q}$ and $A_{\rm Q}$ are, respectively, given by
\begin{align}
\delta r_{\rm Q}:=&\sqrt{|\langle a^2\rangle_N-\langle a\rangle_N^2|}\biggl|_{\langle a\rangle_N=r_{\rm Q}}, \\
\delta A_{\rm Q}:=&\sqrt{|\langle A^2\rangle_N-\langle A\rangle_N^2|}\biggl|_{\langle A\rangle_N=A_{\rm Q}}.
\end{align}
In terms of $x(:=a^{(n-1)/2})$, $\langle a\rangle_N$ and $\langle A\rangle_N$ are written as
\begin{align}
\langle a\rangle_N=&\langle x^{2/(n-1)}\rangle_N, \label{<a>0} \\
\langle A\rangle_N=&V_{n-2}^{(0)}\langle x^{2(n-2)/(n-1)}\rangle_N. \label{<A>0}
\end{align}
In fact, the ``quantum" horizon can in principle be defined using the maximum value of any power of $x$. 
They should all agree in the semi-classical limit, if it exists.

\subsubsection{Euclidean volume of the horizon}
The surface area is proportional to the Bekenstein-Hawking entropy, and hence is a natural geometrical quantity characterizing a black hole.
However, because of the rational powers in the expressions (\ref{<a>0}) and (\ref{<A>0}), it is difficult to obtain $\langle a\rangle_N$ or $\langle A\rangle_N$ in a closed form.
In contrast, the Euclidean volume (or special relativistic volume), which we define as $V_{\rm E}:= V_{n-2}^{(0)} a^{n-1}/(n-1)$, has a computable form in terms of $x$:
\begin{align}
\langle V_{\rm E}\rangle_N=\frac{V_{n-2}^{(0)}}{n-1}\langle x^{2}\rangle_N(t).
\end{align}
In the black-hole spacetime, $V_{\rm E}$ does not represent the volume inside a spacelike hypersurface with constant $r$ because $r$ becomes a timelike coordinate inside the horizon\footnote{It is interesting to note that the Euclidean volume of a stationary black hole is related to the more general concept of a vector volume, defined in Ref.~\cite{ballik2013} by examining the rate of growth of an invariant volume of a spacetime region along a divergence-free vector field.}. 
In order to obtain clear, analytic versions of the main results, we will use the Euclidean volume $V_{\rm E}$ to measure the size of a quantum black hole instead of the surface area $A$ in the present paper.

The Euclidean volume of the classical horizon and its fluctuation are given, respectively, by
\begin{align}
V_{\rm C}:=&\frac{V_{n-2}^{(0)}r_{\rm C}^{n-1}}{n-1}=\frac{2V_{\rm p}l^2\langle M\rangle_N}{(n-2)V_{n-2}^{(0)}l_{\rm p}^2m_{\rm p}},\label{VC-def} \\
\delta V_{\rm C}:=&\frac{2V_{\rm p}l^2\delta \langle M\rangle_N}{(n-2)V_{n-2}^{(0)}l_{\rm p}^2m_{\rm p}}=V_{\rm C}\frac{\delta \langle M\rangle_N}{\langle M\rangle_N},\label{deltaVC-def}
\end{align}
while the Euclidean volume of the quantum horizon and its fluctuation are
\begin{align}
V_{\rm Q}:=&\max_{t\in\mathbb{R}}\langle V_{\rm E}\rangle_N(t)=\frac{V_{n-2}^{(0)}}{n-1}\max_{t\in\mathbb{R}}\langle x^{2}\rangle_N(t),\\
\delta V_{\rm Q}:=&\sqrt{|\langle V_{\rm E}^2\rangle_N-\langle V_{\rm E}\rangle_N^2|}\biggl|_{\langle V_{\rm E}\rangle_N=V_{\rm Q}}.
\end{align}

\subsubsection{Semi-classical limit}
For a standard quantum harmonic oscillator in the stationary state on the whole line $x\in(-\infty,\infty)$, the energy expectation value $\langle E\rangle_N$ and the uncertainty relation between $x$ and its momentum conjugate $p$ are given by
\begin{align}
\langle E\rangle_N=&\hbar\omega\biggl(N+\frac12\biggl),\\
\delta\langle x\rangle_N \delta\langle p\rangle_N=&\hbar\biggl(N+\frac12\biggl),
\end{align}
where $\omega$ is the angular frequency and $N=0,1,2,\cdots$.
The classical limit is formally defined by $\hbar\to 0$ in which canonical pairs of physical quantities become commutable and the spectrum becomes continuous as the energy spacing $\Delta\langle E\rangle_N:=\langle E\rangle_{N+1}-\langle E\rangle_N=\hbar\omega$ converges to zero.

Independently, {and somewhat more practically}, the limit of large $N$ is also often called the semi-classical limit since the relative energy spacing reduces to zero, namely
\begin{align}
\lim_{N\to\infty}\frac{\Delta\langle E\rangle_N}{\langle E\rangle_N}\to 0.
\end{align}
However, this limit does not reproduce a {purely} classical behaviour, since the uncertainty relation still remains valid. 
In fact, the uncertainty in the unit of the Planck constant $\delta\langle x\rangle_N\delta\langle p\rangle_N/\hbar$ is diverging for $N\to \infty$.

We illustrate this point using the quantity $\langle x^2\rangle_N$ because $\delta \langle E\rangle_N=0$ and $\langle x\rangle_N=\langle p\rangle_N=0$ are satisfied in the stationary state.
$\langle x^2\rangle_N$ and its fluctuation are given by
\begin{align}
\langle x^2\rangle_N=&\biggl(N+\frac12\biggl)\ell_0^2,\\
\delta \langle x^2\rangle_N:=&\sqrt{\langle x^4\rangle_N-\langle x^2\rangle_N^2}=\sqrt{\frac{2(N^2+N+1)}{(2N+1)^2}}\langle x^2\rangle_N.
\end{align}
Note that $\ell_0:=\sqrt{\hbar/m\omega}$ is the only length scale in the system, where $m$ is the mass of the oscillator. 
These expressions show 
\begin{align}
\frac{\Delta\langle x^2\rangle_N}{\ell_0^2}=1,\qquad \frac{\delta \langle x^2\rangle_N}{\ell_0^2}=\sqrt{\frac{N^2+N+1}{2}},
\end{align}
where as before $\Delta\langle x^2\rangle_N:= \langle x^2\rangle_{N-1}-\langle x^2\rangle_N$ denotes the spacing between levels. Hence, the quantum fluctuation $\delta \langle x^2\rangle_N$ increases as $N$ increases and cannot be negligible in comparison with the scale $\ell_0^2$ even for $N=0$.
Alternatively, one may use the following ratios to evaluate the fluctuations:
\begin{align}
\frac{\Delta\langle x^2\rangle_N}{\langle x^2\rangle_N}=&\frac{2}{2N+1},\\
\frac{\delta \langle x^2\rangle_N}{\langle x^2\rangle_N}=&\sqrt{\frac{2(N^2+N+1)}{(2N+1)^2}}.
\end{align}
While the relative spacing $\Delta\langle x^2\rangle_N/\langle x^2\rangle_N$ converges to zero for large $N$, the relative fluctuation $\delta\langle x^2\rangle_N/\langle x^2\rangle_N$ reduces to $1/\sqrt{2}\simeq 0.7071$ for $N\to\infty$. {Of course, for the simple harmonic oscillator on the whole line, the semi-classical behaviour is verified  by observing, for example, the shape of the probability amplitude for the position operator $x$. In the large-$N$ limit the quantum probability of finding the particle in an finite segment of the $x$-axis coincides with the corresponding classical property obtained by calculating how much time the oscillating particle spends in that region. In both cases, the probability of finding the particle at the turning points, for example, is greatest, whereas the probability of finding it near the equilibrium point is small.}

In the context of quantum gravity, one expects classical behaviour for black holes that are sufficiently large.
For dynamical quantum black holes, one may therefore define the ``semi-classical limit"  by  requiring that the quantum fluctuations to be negligible--namely that $\delta \langle X\rangle/\langle X\rangle\ll 1$ for any physical quantity $X$ associated with the black hole.
In such a limit, both $A_{\rm Q}/A_{\rm C}\to 1$ and $V_{\rm Q}/V_{\rm C}\to 1$ are also expected to hold, so that the classical and quantum horizons coincide.
One of the main purposes of the present paper is to identify such a limit for the dynamical quantum black holes described by exact wave functions.\\

\section{{Exact time-dependent wave functions for toroidal black holes}}
\label{sec:wave}
In the previous section, we  summarized our main results for stationary quantum black holes. 
In this section, we will present exact time-dependent wave functions describing quantum black holes in non-stationary states.
We focus on the case of toroidal black holes ($k=0$) whose Schr\" odinger equation (\ref{Schro-eq}) reduces to that of the simple harmonic oscillator.
By the coordinate transformations $x=\sqrt{(n-1)^2\hbar \kappa_n^2/4(n-2)V_{n-2}^{(0)}}{\bar x}$,  the Schr\" odinger equation (\ref{Schro-eq}) with $k=0$ is cast into the following standard form:
\begin{align}
\biggl(-\frac12\frac{\partial^2}{\partial{\bar x}^2}+\frac12\omega^2{\bar x}^2\biggl)\Psi=&i\frac{\partial\Psi}{\partial t},\label{tosolve}
\end{align}  
where
\begin{align}
\omega:=\frac{n-1}{2l}.\label{def-omega}
\end{align}

As anticipated, this equation is precisely that of a quantum harmonic oscillator on the half line ${\bar x}\in [0,\infty)$. In a previous paper~\cite{km2014}, we obtained the exact stationary-state mass spectrum (\ref{EN:toroidal}) for the Dirichlet or Neumann boundary condition at ${\bar x}=0$.
We now take advantage of the fact that for this system exact time-dependent solutions of the Schr\" odinger equation are also available.

\subsection{Time-dependent wave function on the whole line: A review}
\label{sec:harmonic-review}

There is a six-parameter family of exact solutions to the Schr\" odinger equation (\ref{tosolve})~\cite{lsv2013}~\footnote{We have done coordinate transformations and reparametrization from the original expressions in Ref.~\cite{lsv2013}.}:
\begin{align}
\Psi({\bar x},t)=&\Psi_N({\bar x},t) \nonumber \\
:=&\frac{e^{i\left(\alpha(t){\bar x}^2+\delta(t){\bar x}+\kappa(t)\right)+i(2N+1)\gamma(t)}}{\sqrt{2^NN!\mu(t)\sqrt{\pi}}} e^{-(\beta(t){\bar x}+\varepsilon(t))^2/2}H_N(\beta(t){\bar x}+\varepsilon(t)). \label{six-solution}
\end{align}
In the above, $H_N$ are the Hermite polynomials with integer $N(=0,1,2,\cdots)$ and 
\begin{align}
\mu(t)=&\mu_0\sqrt{{\bar\beta}_0^4\sin^2\omega t/\omega ^2+(2\alpha_0\sin \omega t+\cos \omega t)^2}, \label{six-sol1}\\
\alpha(t)=&\frac{\omega \alpha_0\cos 2\omega t+(\sin 2\omega t/\omega )({\bar\beta}_0^4+4\omega ^2\alpha_0^2-\omega ^2)/4}{{\bar\beta}_0^4\sin^2\omega t/\omega ^2+(2\alpha_0\sin \omega t+\cos \omega t)^2},\\
\beta(t)=&\frac{{\bar\beta}_0}{\sqrt{{\bar\beta}_0^4\sin^2\omega t/\omega ^2+(2\alpha_0\sin \omega t+\cos \omega t)^2}},\\
\gamma(t)=&\gamma_0-\frac12\arctan\biggl(\frac{{\bar\beta}_0^2\sin \omega t/\omega }{2\alpha_0\sin \omega t+\cos \omega t}\biggl),\\
\delta(t)=&\frac{{\bar\delta}_0(2\alpha_0\sin \omega t+\cos \omega t)+\varepsilon_0{\bar\beta}_0^3\sin \omega t/\omega }{{\bar\beta}_0^4\sin^2\omega t/\omega ^2+(2\alpha_0\sin \omega t+\cos \omega t)^2},\\
\varepsilon(t)=&\frac{\varepsilon_0(2\alpha_0\sin \omega t+\cos \omega t)-{\bar\beta}_0{\bar\delta}_0\sin \omega t/\omega }{\sqrt{{\bar\beta}_0^4\sin^2\omega t/\omega ^2+(2\alpha_0\sin \omega t+\cos \omega t)^2}},\\
\kappa(t)=&\kappa_0+\frac{\sin^2 \omega t}{\omega ^2}\frac{\varepsilon_0{\bar\beta}_0^2(\omega \alpha_0\varepsilon_0-{\bar\beta}_0{\bar\delta}_0)-\omega \alpha_0{\bar\delta}_0^2}{{\bar\beta}_0^4\sin^2\omega t/\omega ^2+(2\alpha_0\sin \omega t+\cos \omega t)^2} \nonumber \\
&+\frac14\frac{\sin 2\omega t}{\omega }\frac{\varepsilon_0^2{\bar\beta}_0^2-{\bar\delta}_0^2}{{\bar\beta}_0^4\sin^2\omega t/\omega ^2+(2\alpha_0\sin \omega t+\cos \omega t)^2},\label{six-sol2}
\end{align}
where $\mu_0,{\alpha}_0,{\bar \beta}_0,\gamma_0,{\bar\delta}_0,\kappa_0,\varepsilon_0$ are constants.
Among them, $\gamma_0$ and $\kappa_0$ are phase constants. 
In order to verify that the solution (\ref{six-solution}) solves the Schr\" odinger equation (\ref{tosolve}), the following relations are useful:
\begin{align}
& {\dot \mu}=2\mu \alpha,\quad {\dot \alpha}=\frac12(\beta^4-4\alpha^2-\omega^2),\label{t-derivative1} \\
& {\dot \beta}=-2\beta\alpha,\quad {\dot \gamma}=-\frac12\beta^2,\quad  {\dot \delta}=\beta^3\varepsilon-2\alpha\delta, \label{t-derivative4}\\
& {\dot \varepsilon}=-\beta \delta,\quad {\dot \kappa}=\frac12(\beta^2\varepsilon^2-\delta^2),\label{t-derivative7}
\end{align}
where a dot denotes differentiation with respect to $t$.

Using the orthogonality condition (\ref{Hermite-orthogonal}) of the Hermite polynomials, we compute the squared norm of $\Psi_N$ on the whole line:
\begin{align}
\langle \Psi_N|\Psi_N \rangle=\int_{-\infty}^\infty|\Psi_N({\bar x},t)|^2\D {\bar x}=\frac{1}{\mu(t) \beta(t)}=\frac{1}{\mu_0{\bar \beta}_0}.
\end{align}
Since normalization requires $\mu_0{\bar\beta}_0=1$, the number of independent parameters is six, of which two ($\kappa_0$ and $\gamma_0$) are pure phase. The physical properties of the solution are therefore characterized by four continuous parameters ${\alpha}_0$, ${\bar \beta}_0$, ${\bar\delta}_0$, and $\varepsilon_0$ in addition to the quantum number $N$.

Within this class of solutions, stationary states are realized as special cases for which ${\alpha}_0={\bar\delta}_0=\varepsilon_0=0$ and ${\bar \beta}_0=\sqrt{\omega}$. In this case we have
\begin{align}
\mu(t)=&\mu_0,\quad \alpha(t)=0,\quad \beta(t)=\sqrt{\omega},\\
 \gamma(t)=&\gamma_0-\frac12\omega t, \quad \delta(t)=\varepsilon(t)=0,\quad \kappa(t)=\kappa_0.
\end{align}

We will use this time-dependent solution (\ref{six-solution}) to construct exact wave functions on the half line, but first we will review some properties of the solution on the whole line.
The details of computations are presented in Appendix~\ref{app:wholeline}.

\subsubsection{Energy expectation value and uncertainty relation}
The energy expectation value is given by 
\begin{align}
\langle E \rangle_N=&\frac{\langle \Psi_N|{\hat H}\Psi_N \rangle}{\langle \Psi_N|\Psi_N \rangle}=\biggl(N+\frac12\biggl)\hbar\Omega+\frac12\hbar\eta, \label{Energy-wholeline}
\end{align}
where ${\hat H}=i\hbar\partial/\partial t$ is the Hamiltonian operator and 
\begin{align}
\Omega:=&\frac{{\bar\beta}_0^4+4\alpha_0^2\omega^2+\omega^2}{2{\bar\beta}_0^2}, \label{Omega-def}\\
\eta:=&\frac{(2\alpha_0\varepsilon_0\omega-{\bar\beta}_0{\bar \delta}_0)^2+\varepsilon_0^2\omega^2}{{\bar\beta}_0^{2}}.\label{eta-def}
\end{align}
Hence, the expectation value of the energy is time independent and equally spaced, with $\Delta \langle E\rangle_N:=\langle E\rangle_{N-1}-\langle E\rangle_N=\hbar\Omega$.
On the other hand, the fluctuation of the energy $\delta \langle E\rangle_N:=\sqrt{\langle E^2\rangle_N-\langle E\rangle_N^2}$ is in general given by a complicated expression that is time dependent.
However,  there is a subclass of states for which both the energy expectation value and the energy fluctuation are constant. These are  the so-called ``shape-preserving states'', which as the name suggests are time-dependent states whose probability amplitudes move without deformation. They include as a special case the usual coherent state of the simple harmonic oscillator and will be discussed in full in Sec.~\ref{sec:exact2}.

It turns out that the energy expectation value (\ref{Energy-wholeline}) can in general be written as
\begin{align}
\langle E \rangle_N=&\biggl(N+\frac12\biggl)\hbar\Omega+\frac{\hbar}{2}\biggl(\omega^2\langle {\bar x}\rangle_N^2+\hbar^{-2}\langle p\rangle_N^2\biggl),
\end{align}
where $\langle {\bar x}\rangle_N$ and $\langle p\rangle_N$ are expectation values of the position ${\bar x}$ and the momentum ${\hat p}=-i\hbar\partial/\partial{\bar x}$, respectively, given by 
\begin{align}
\langle {\bar x}\rangle_N=&-\frac{\varepsilon}{\beta} \nonumber \\
=&\frac{1}{\omega}\biggl\{\biggl({\bar \delta}_0-\frac{2\alpha_0\varepsilon_0\omega}{{\bar \beta}_0}\biggl)\sin\omega t-\frac{\varepsilon_0\omega}{{\bar \beta}_0}\cos\omega t\biggl\},\label{exp-x-full}\\
\langle p\rangle_N=&\hbar\biggl(\delta-\frac{2\alpha\varepsilon}{\beta}\biggl) \nonumber \\
=&\hbar\biggl\{\biggl({\bar \delta}_0-\frac{2\alpha_0\varepsilon_0\omega}{{\bar \beta}_0}\biggl)\cos\omega t+\frac{\varepsilon_0\omega}{{\bar \beta}_0}\sin\omega t\biggl\}.\label{exp-p-full}
\end{align}
Thus, the constant $\eta$ represents the sum of the kinetic and potential energies.

The fluctuations of the position and momentum are given by  
\begin{align}
\delta \langle {\bar x}\rangle_N:=&\sqrt{\langle {\bar x}^2\rangle_N-\langle {\bar x}\rangle_N^2}=\sqrt{\frac{2N+1}{2\beta^2}} \nonumber \\
=&\sqrt{\frac{(2N+1)}{2{\bar\beta}_0^2}\biggl\{\frac{{\bar\beta}_0^4}{\omega^2}\sin^2\omega t+(2\alpha_0\sin \omega t+\cos \omega t)^2\biggl\}},\\
\delta \langle p\rangle_N:=&\sqrt{\langle p^2\rangle_N-\langle p\rangle_N^2}=\sqrt{\frac{\hbar^2(2N+1)}{2\beta^2}(4\alpha^2+\beta^{4})} \nonumber \\
=&\sqrt{\frac{\hbar^2(2N+1)}{2{\bar\beta}_0^2}\biggl\{{\bar\beta}_0^4\cos^2\omega t+\omega^2(\sin\omega t-2\alpha_0\cos\omega t)^2\biggl\}}.
\end{align}
Hence, the uncertainty relation is given by 
\begin{align}
&\delta \langle {\bar x}\rangle_N\delta \langle p\rangle_N \nonumber \\
&~~=\frac{\hbar(2N+1)}{4{\bar\beta}_0^2\omega}\biggl\{(4\alpha_0^2\omega^2+{\bar\beta}_0^4+\omega^2)^2 -\biggl((4\alpha_0^2\omega^2+{\bar\beta}_0^4-\omega^2)\cos2\omega t-4\alpha_0\omega^2\sin2\omega t\biggl)^2\biggl\}^{1/2}, \label{uncertainty-full}
\end{align}
where the expression inside the square root is positive definite.

\subsubsection{Shape-preserving state}
\label{sec:ShapePreserving}
The  shape-preserving states are realized for $\alpha_0=0$ and ${\bar\beta}_0=\sqrt{\omega}$, in which case one has:
\begin{align}
\mu(t)=&\mu_0,\quad \alpha(t)=0,\quad \beta(t)=\sqrt{\omega},\quad \gamma(t)=\gamma_0-\frac12\omega t,
\label{eq:SPmu}\\
\delta(t)=&{\bar\delta}_0\cos \omega t+\varepsilon_0\sqrt{\omega}\sin \omega t=\sqrt{\omega\zeta}\sin (\omega t+\theta_1),\label{eq:SPdelta}\\
\varepsilon(t)=&\varepsilon_0\cos \omega t-{\bar\delta}_0\sin \omega t/\sqrt{\omega}=\sqrt{\zeta}\sin (\omega t+\theta_2),\label{eq:SPepsilon}\\
\kappa(t)=&\kappa_0-\frac{\varepsilon_0{\bar\delta}_0\sin^2 \omega t}{\sqrt{\omega}} +\frac{(\varepsilon_0^2\omega-{\bar\delta}_0^2)\sin 2\omega t}{4\omega },
\label{eq:SPkappa}
\end{align}
where $\theta_1:=\arctan({\bar\delta}_0/\varepsilon_0\sqrt{\omega})$, $\theta_2:=\arctan(-\varepsilon_0\sqrt{\omega}/{\bar\delta}_0)$, and 
\begin{align}
\zeta:=\varepsilon^2+\frac{\delta^2}{\beta^2}=\varepsilon_0^2+\frac{{\bar\delta}_0^2}{\omega}. \label{zeta-def}
\end{align}
$\Omega=\omega$ and $\eta=\omega\zeta$ hold in this shape-preserving state and $\langle E\rangle_N$, $\langle {\bar x}\rangle_N$, and $\langle p\rangle_N$ reduce to
\begin{align}
\langle E\rangle_N=&\biggl(N+\frac12\biggl)\hbar\omega+\frac12\hbar\omega\zeta,\label{energy-shapepreserving}\\
\langle {\bar x}\rangle_N=&-\frac{\varepsilon}{\sqrt{\omega}},\quad \langle p\rangle_N=\hbar\delta.
\end{align}
A characteristic property of the shape-preserving states is that the uncertainty relation (\ref{uncertainty-full}) becomes constant:
\begin{align}
\delta \langle {\bar x}\rangle_N\delta \langle p\rangle_N=\hbar\biggl(N+\frac12\biggl). 
\end{align}

For shape-preserving states, the relative fluctuation of $\langle {\bar x}\rangle_N$ is given by 
\begin{align}
\frac{\delta \langle {\bar x}\rangle_N}{\langle {\bar x}\rangle_N}=&\frac{\sqrt{\langle {\bar x}^2\rangle_N-\langle {\bar x}\rangle_N^2}}{\langle {\bar x}\rangle_N}=-\frac{1}{\varepsilon}\sqrt{\frac{2N+1}{2}}.
\end{align}
The absolute value of this quantity diverges for large $N$ as 
\begin{align}
\lim_{N\to \infty}\biggl|\frac{\delta \langle {\bar x}\rangle_N}{\langle {\bar x}\rangle_N}\biggl|\simeq |\varepsilon|^{-1}N^{1/2}\to \infty.
\end{align}
Because the range of $\varepsilon(t)$ is $-\sqrt{\zeta}\le \varepsilon\le \sqrt{\zeta}$, we obtain
\begin{align}
\min_{t\in\mathbb{R}}\biggl|\frac{\delta \langle {\bar x}\rangle_N}{\langle {\bar x}\rangle_N}\biggl|=\sqrt{\frac{2N+1}{2\zeta}},\qquad \max_{t\in\mathbb{R}}\biggl|\frac{\delta \langle {\bar x}\rangle_N}{\langle {\bar x}\rangle_N}\biggl|=\infty
\end{align}
for a given value of $N$, where the minimum and maximum are realized when $|\langle {\bar x}\rangle_N|=\max_{t\in\mathbb{R}}|\langle {\bar x}\rangle_N|=\sqrt{\zeta/\omega}$ and $\langle {\bar x}\rangle_N=0$ hold, respectively.
The minimum of the relative fluctuation reduces to zero for $\zeta\to \infty$ as
\begin{align}
\lim_{\zeta\to \infty}\min_{t\in\mathbb{R}}\biggl|\frac{\delta \langle {\bar x}\rangle_N}{\langle {\bar x}\rangle_N}\biggl|\simeq \sqrt{\frac{N}{\zeta}}\to 0.
\end{align}
Actually, the limit $\zeta\to \infty$ is unphysical since it gives infinite energy; however, the minimum of the relative fluctuation becomes very small in the situation of $\zeta\gg N$.
Equation~(\ref{energy-shapepreserving}) shows that, for $\zeta\gg N$, the energy of the shape-preserving state is much larger than the energy in the stationary state.

\begin{figure}[htbp]
\begin{center}
\includegraphics[width=0.7\linewidth]{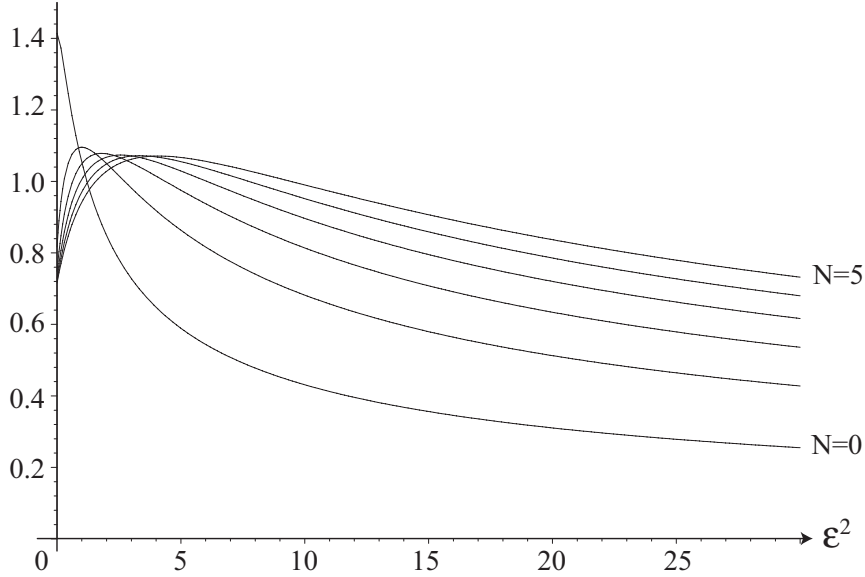}
\caption{\label{deltax2-coherent} $\delta \langle {\bar x}^2\rangle_N/\langle {\bar x}^2\rangle_N$ as a function of $\varepsilon^2$ for the shape-preserving state with $N=0,1,\cdots,5$.}
\end{center}
\end{figure}
For later reference, we also present the relative fluctuation of $\langle {\bar x}^2\rangle_N$:
\begin{align}
\frac{\delta \langle {\bar x}^2\rangle_N}{\langle {\bar x}^2\rangle_N}=&\frac{\sqrt{\langle {\bar x}^4\rangle_N-\langle {\bar x}^2\rangle_N^2}}{\langle {\bar x}^2\rangle_N} \nonumber \\
=&\frac{\sqrt{2(N^2+N+1)+8(2N+1)\varepsilon^2}}{(2N+1)+2\varepsilon^2}.\label{deltax2-shape}
\end{align}
This quantity has a finite limit for $N\to \infty$ as
\begin{align}
\lim_{N\to \infty}\frac{\delta \langle {\bar x}^2\rangle_N}{\langle {\bar x}^2\rangle_N}=\frac{1}{\sqrt{2}}.
\end{align}
Figure~\ref{deltax2-coherent} shows the shape of $\delta \langle {\bar x}^2\rangle_N/\langle {\bar x}^2\rangle_N$ as a function of $\varepsilon^2$ for $N=0,1,\cdots,5$.
It has a maximum at $\varepsilon^2=3N(N+1)/2(2N+1)$.
The relative fluctuation at $\varepsilon=\pm\sqrt{\zeta}$ is given by 
\begin{align}
\frac{\delta \langle {\bar x}^2\rangle_N}{\langle {\bar x}^2\rangle_N}\biggl|_{\varepsilon^2=\zeta}=&\frac{\sqrt{2(N^2+N+1)+8(2N+1)\zeta}}{(2N+1)+2\zeta},
\end{align}
which reduces to zero for $\zeta\to \infty$ as
\begin{align}
\lim_{\zeta\to \infty}\frac{\delta \langle {\bar x}^2\rangle_N}{\langle {\bar x}^2\rangle_N}\biggl|_{t=t_{\rm max}}\simeq &\frac{\sqrt{2(2N+1)}}{\zeta^{1/2}}\to 0.
\end{align}

\subsection{Time-dependent wave function on the half line}

The wave function (\ref{six-solution}) is an exact solution of the Schr\" odinger equation (\ref{tosolve}) on the whole line ${\bar x}\in(-\infty,\infty)$.
By symmetrizing or anti-symmetrizing this  solution about ${\bar x}=0$, one can construct exact wave functions on the half line ${\bar x}\in[0,\infty)$ satisfying Neumann or Dirichlet boundary conditions, respectively, at the origin. 
This yields
\begin{align}
\Psi^{(\pm)}_{N}({\bar x},t):=&\frac12\biggl(\Psi_N({\bar x},t)\pm \Psi_N(-{\bar x},t)\biggl) \nonumber \\
=&\frac{e^{i(\alpha{\bar x}^2+\kappa)+i(2N+1)\gamma}}{2\sqrt{2^NN!\mu\sqrt{\pi}}}\biggl\{e^{i\delta{\bar x}} e^{-(\beta{\bar x}+\varepsilon)^2/2}H_N(\beta{\bar x}+\varepsilon)\pm e^{-i\delta{\bar x}}e^{-(\beta{\bar x}-\varepsilon)^2/2}H_N(-\beta{\bar x}+\varepsilon)\biggl\}, \label{SolHalfLine}
\end{align}
which satisfies Neumann (Dirichlet) boundary conditions at ${\bar x}=0$ for the upper (lower) sign.

The square of the wave function is an even function given by 
\begin{align}
|\Psi^{(\pm)}_{N}|^2=&\frac{1}{2^{N+2}N!\mu\sqrt{\pi}}\biggl\{e^{-(\beta{\bar x}+\varepsilon)^2}H_N(\beta{\bar x}+\varepsilon)^2 \nonumber \\
&+e^{-(\beta{\bar x}-\varepsilon)^2}H_N(-\beta{\bar x}+\varepsilon)^2\pm 2\cos(2\delta{\bar x}) e^{-\beta^2{\bar x}^2-\varepsilon^2}H_N(\beta{\bar x}+\varepsilon)H_N(-\beta{\bar x}+\varepsilon) \biggl\}.
\label{SquareNormGeneral}
\end{align}
While the integral of the first two terms are easily evaluated to yield
\begin{align}
&\int_{0}^\infty e^{-(\beta{\bar x}+\varepsilon)^2}H_N(\beta{\bar x}+\varepsilon)^2 \D {\bar x} \nonumber \\
=&\int_{0}^\infty e^{-(\beta{\bar x}-\varepsilon)^2}H_N(-\beta{\bar x}+\varepsilon)^2  \D {\bar x}=\frac{\sqrt{\pi}2^{N-1}N!}{\beta},
\end{align}
it is difficult to compute the last term for general $N$.
Nevertheless, using the following integral and its derivatives with respect to $q$,
\begin{align}
\int_{0}^\infty \cos (2qy) e^{-y^2}\D y=&\frac12\int_{-\infty}^\infty \cos (2qy)e^{-y^2}\D y \nonumber \\
=&\frac{\sqrt{\pi}}{2}e^{-q^2},
\end{align}
one can compute the squared norm for each value of $N$.

Equation~(\ref{SolHalfLine}) provides two families of dynamical solutions to the harmonic oscillator Schr\" odinger equation on the half line, one satisfying Neumann boundary conditions at the origin, the other satisfying Dirichlet boundary conditions. Each family retains all the parameters of the solutions on the whole line. For the reasons given before, the  physical parameters are $\varepsilon_0$, ${\bar \delta}_0$, $\alpha_0$ , ${\bar \beta}_0$ and the integer $N$. We have omitted $\mu_0$ since it is determined in terms of $\varepsilon_0$ and ${\bar \beta}_0$ via the normalization condition (cf. Eq.~(\ref{SquareNormGeneral})).  We emphasize that the two families live in distinct Hilbert spaces in which the Hamiltonian is self-adjoint.

Given that there is such a  large parameter space of solutions, it is difficult to analyze the dynamical behaviour in full generality. In the following, we will therefore focus on two different subclasses of solutions of particular interest. The subclasses retain two distinct pairs of parameters and are characterized by different dynamical behaviour. 

\subsubsection{Wave function I}

The first class of wave functions on the half line is obtained from  (\ref{SolHalfLine}) by setting  ${\bar\delta}_0=\varepsilon_0=0$ (and hence $\delta(t)=\varepsilon(t)=0$). It is interesting to note that making this substitution directly into the wave function (\ref{six-solution}) also yields Neumann or Dirichlet boundary conditions, without the need to (anti-)symmetrize first~\footnote{If one chooses to (anti-)symmetrize first, then the anti-symmetric solutions vanish identically for even $N$, while the symmetric solutions vanish identically for odd $N$. The remaining solutions are precisely those given above.}. One obtains
\begin{align}
\Psi_{{\rm I}(N)}({\bar x},t):=&\frac{e^{i(\alpha(t){\bar x}^2+\kappa(t))+i(2N+1)\gamma(t)}}{\sqrt{2^NN!\mu(t)\sqrt{\pi}}}e^{-\beta(t)^2{\bar x}^2/2}H_N(\beta(t){\bar x}), \label{sol-first}
\end{align}
where 
\begin{align}
\mu(t)=&\mu_0\sqrt{{\bar\beta}_0^4\sin^2\omega t/\omega ^2+(2\alpha_0\sin \omega t+\cos \omega t)^2}, \\
\alpha(t)=&\frac{\omega \alpha_0\cos 2\omega t+(\sin 2\omega t/\omega )({\bar\beta}_0^4+4\omega ^2\alpha_0^2-\omega ^2)/4}{{\bar\beta}_0^4\sin^2\omega t/\omega ^2+(2\alpha_0\sin \omega t+\cos \omega t)^2},\\
\beta(t)=&\frac{{\bar\beta}_0}{\sqrt{{\bar\beta}_0^4\sin^2\omega t/\omega ^2+(2\alpha_0\sin \omega t+\cos \omega t)^2}},\\
\gamma(t)=&\gamma_0-\frac12\arctan\biggl(\frac{{\bar\beta}_0^2\sin \omega t/\omega }{2\alpha_0\sin \omega t+\cos \omega t}\biggl),\\
\kappa(t)=&\kappa_0.
\end{align}
Neumann (Dirichlet) boundary conditions are satisfied for even (odd) $N$.  In this case, the normalization can be carried out explicitly:
\begin{align}
\langle \Psi_{{\rm I}(N)}|\Psi_{{\rm I}(N)}\rangle=&\int_{0}^\infty|\Psi_{{\rm I}(N)}({\bar x},t)|^2\D {\bar x} \nonumber \\
=&\frac{1}{2\mu(t)\beta(t)}=\frac{1}{2\mu_0{\bar \beta}_0}. \label{s-norm-half}
\end{align}
As claimed previously, this class of solutions has two independent continuous physical parameters $\alpha_0$ and ${\bar \beta}_0$ in addition to the quantum number $N$.  The parameters $\gamma_0$ and $\kappa_0$ are, as before, pure phase, while the parameter $\mu_0$ is determined by normalization of $\Psi_{{\rm I}(N)}$.
Stationary states on the half line are realized for all $N$ when $\alpha_0=0$ and ${\bar\beta}_0=\sqrt{\omega}$.

\subsubsection{Wave function II}
The second class of exact wave functions $\Psi=\Psi^{(\pm)}_{{\rm II}(N)}({\bar x},t)$ on the half line is given by setting $\alpha_0=0$ and $\beta_0=\sqrt{\omega}$ into the (anti-)symmetrized wave functions (\ref{SolHalfLine}). One thereby obtains subclasses of (anti-)symmetric states $\Psi^{(\pm)}_{{\rm II}(N)}$. These are the analogues from the state
 on the half line of the shape-preserving solutions described in Sec.~\ref{sec:ShapePreserving}. They can also be obtained directly by (anti-)symmetrizing the shape-preserving solutions on the whole line. 

The wave function $\Psi^{(\pm)}_{{\rm II}(N)}({\bar x},t)$ is given by Eq.~(\ref{SolHalfLine}) with the functions (\ref{eq:SPmu})--(\ref{eq:SPkappa}).  It is useful to exhibit the first six squared norms:
\begin{align}
\int_0^\infty|\Psi^{(\pm)}_{{\rm II}(0)}|^2\D {\bar x}=&\frac{1}{4\mu\beta}\biggl\{1\pm e^{-\zeta}\biggl\},\\
\int_0^\infty|\Psi^{(\pm)}_{{\rm II}(1)}|^2\D {\bar x}=&\frac{1}{4\mu\beta}\biggl\{1\pm \biggl(2\zeta-1\biggl)e^{-\zeta}\biggl\},\\
\int_0^\infty|\Psi^{(\pm)}_{{\rm II}(2)}|^2\D {\bar x}=&\frac{1}{4\mu\beta}\biggl\{1\pm \biggl(2\zeta^2-4\zeta+1\biggl)e^{-\zeta}\biggl\},\\
\int_0^\infty|\Psi^{(\pm)}_{{\rm II}(3)}|^2\D {\bar x}=&\frac{1}{12\mu\beta}\biggl\{3\pm \biggl(4\zeta^3-18\zeta^2+18\zeta-3\biggl)e^{-\zeta}\biggl\},\\
\int_0^\infty|\Psi^{(\pm)}_{{\rm II}(4)}|^2\D {\bar x}=&\frac{1}{12\mu\beta}\biggl\{3\pm \biggl(2\zeta^4-16\zeta^3+36\zeta^2-24\zeta+3\biggl)e^{-\zeta}\biggl\},\\
\int_0^\infty|\Psi^{(\pm)}_{{\rm II}(5)}|^2\D {\bar x}=&\frac{1}{60\mu\beta}\biggl\{15\pm \biggl(4\zeta^5-50\zeta^4+200\zeta^3-300\zeta^2+150\zeta-15\biggl)e^{-\zeta}\biggl\},
\end{align}
where
\begin{align}
\zeta:=\varepsilon(t)^2+\frac{\delta(t)^2}{\beta(t)^2}=\varepsilon_0^2+\frac{{\bar\delta}_0^2}{{\bar\beta}_0^2}.
\label{eq:zeta}
\end{align}
It is interesting to note that the physical properties of these solutions depend only on the time-independent combination of $\varepsilon(t)$ and $\delta(t)$ given by $\zeta$, which measures the difference from the stationary state (${\bar \delta}_0=0$ and ${\varepsilon}_0=0$) or a vanishing wave function $\Psi^{(\pm)}_{{\rm II}(N)}\equiv 0$.
The latter is realized for $\zeta=0$ in the case of the upper sign with odd $N$ or the lower sign with even $N$.

\begin{figure*}[htbp]
\begin{center}
\includegraphics[width=1.0\linewidth]{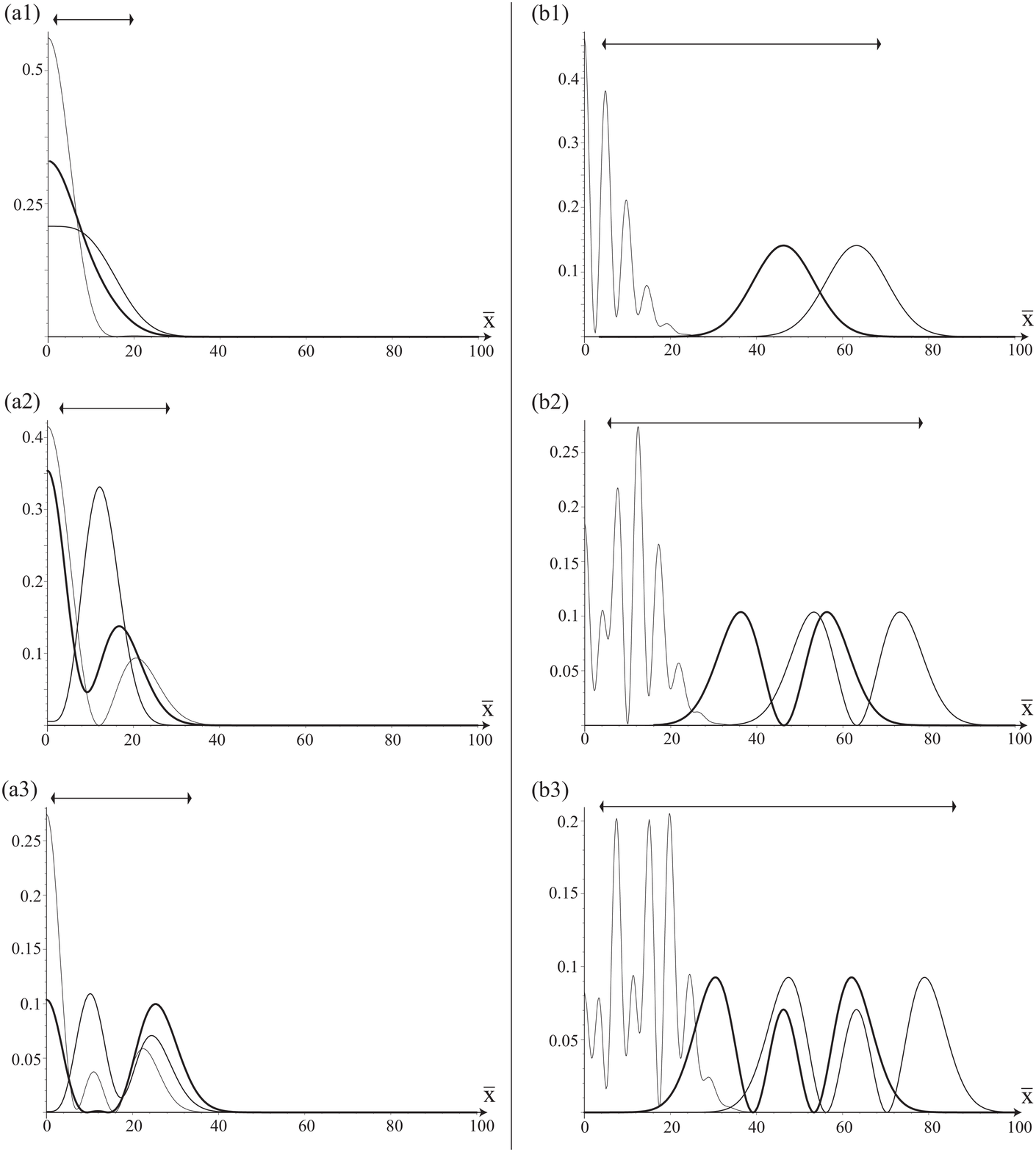}
\caption{\label{fig-wave2(all-)} Three snapshots during the evolution of the unnormalized squared norm $|\Psi^{(+)}_{{\rm II}(N)}({\bar x},t)|^2$ with $\mu_0=1$, $\omega=1/100$, and ${\bar\delta}_0=0$.
The left row corresponds to $\varepsilon_0^2=1$ ($\zeta=1$) with (a1) $N=0$, (a2) $N=1$, and (a3) $N=2$.
The right row corresponds to $\varepsilon_0^2=40$ ($\zeta=40$) with (b1) $N=0$, (b2) $N=1$, and (b3) $N=2$.
}
\end{center}
\end{figure*}

\begin{figure*}[htbp]
\begin{center}
\includegraphics[width=1.0\linewidth]{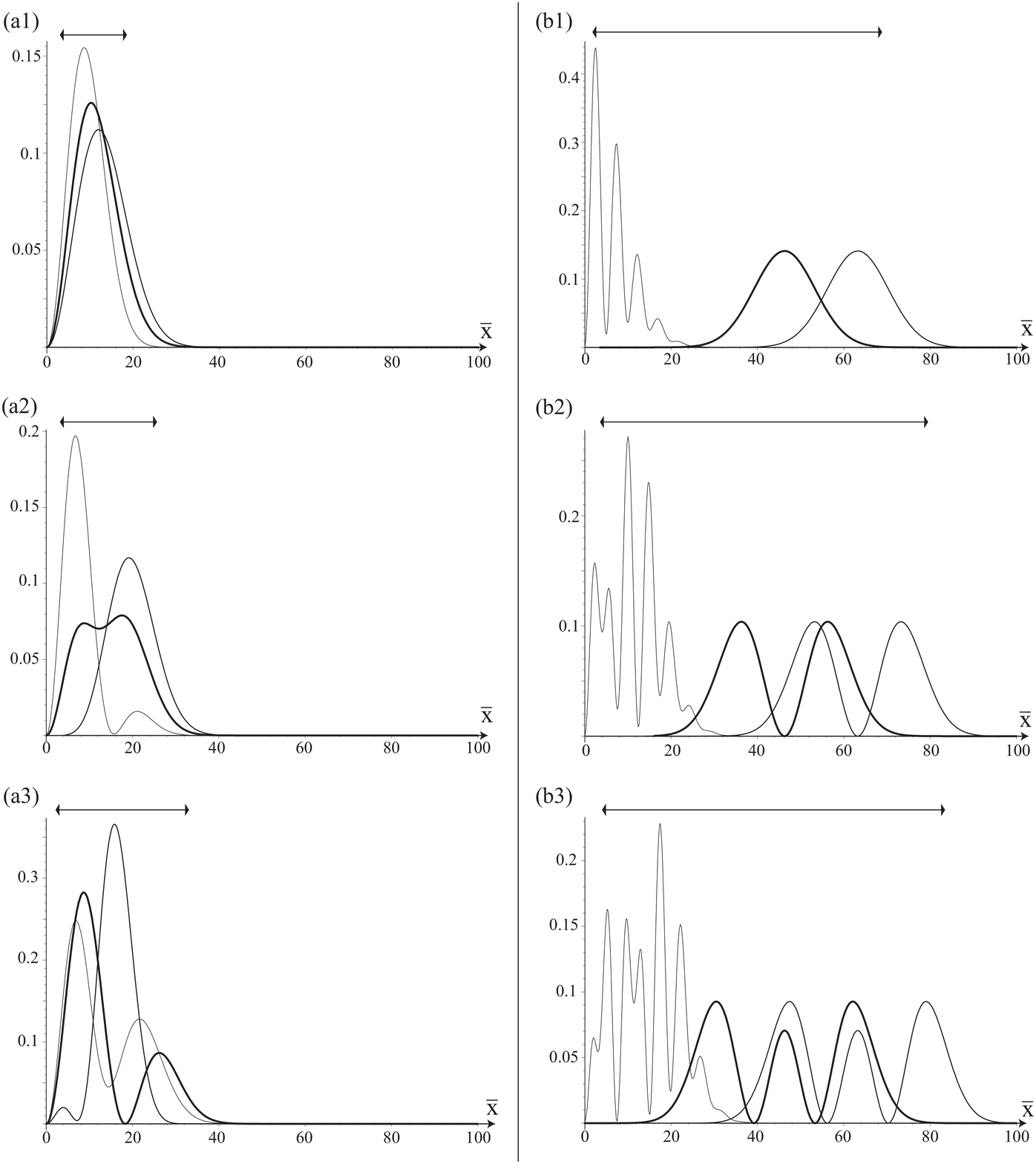}
\caption{\label{fig-wave2(all+)} Three snapshots during the evolution of the unnormalized squared norm $|\Psi^{(-)}_{{\rm II}(N)}({\bar x},t)|^2$ with $\mu_0=1$, $\omega=1/100$, and ${\bar\delta}_0=0$.
The left row corresponds to $\varepsilon_0^2=1$ ($\zeta=1$) with (a1) $N=0$, (a2) $N=1$, and (a3) $N=2$.
The right row corresponds to $\varepsilon_0^2=40$ ($\zeta=40$) with (b1) $N=0$, (b2) $N=1$, and (b3) $N=2$.
}
\end{center}
\end{figure*}

The evolutions of the (unnormalized) squared norm $|\Psi^{(+)}_{{\rm II}(N)}|^2$ and $|\Psi^{(-)}_{{\rm II}(N)}|^2$ with $N=0,1,2$ are drawn in Figs.~\ref{fig-wave2(all-)} and \ref{fig-wave2(all+)}, respectively.
They show that, while the evolution is rather chaotic for $\zeta=1$, the wave packet is shape preserving when it is away from the origin for $\zeta=40$.

\section{Quantum toroidal black hole: Wave function I}
\label{sec:exact1}

In this section, we will study quantum toroidal black holes described by the wave function (\ref{sol-first}). Hereafter we assume ${\bar \beta}_0\ge 0$ (hence $\beta(t)\ge 0$) without loss of generality.
The details of computations are presented in Appendix~\ref{app:halfline}.

\subsection{Dynamical singularity resolution}

\begin{figure*}[htbp]
\begin{center}
\includegraphics[width=1.0\linewidth]{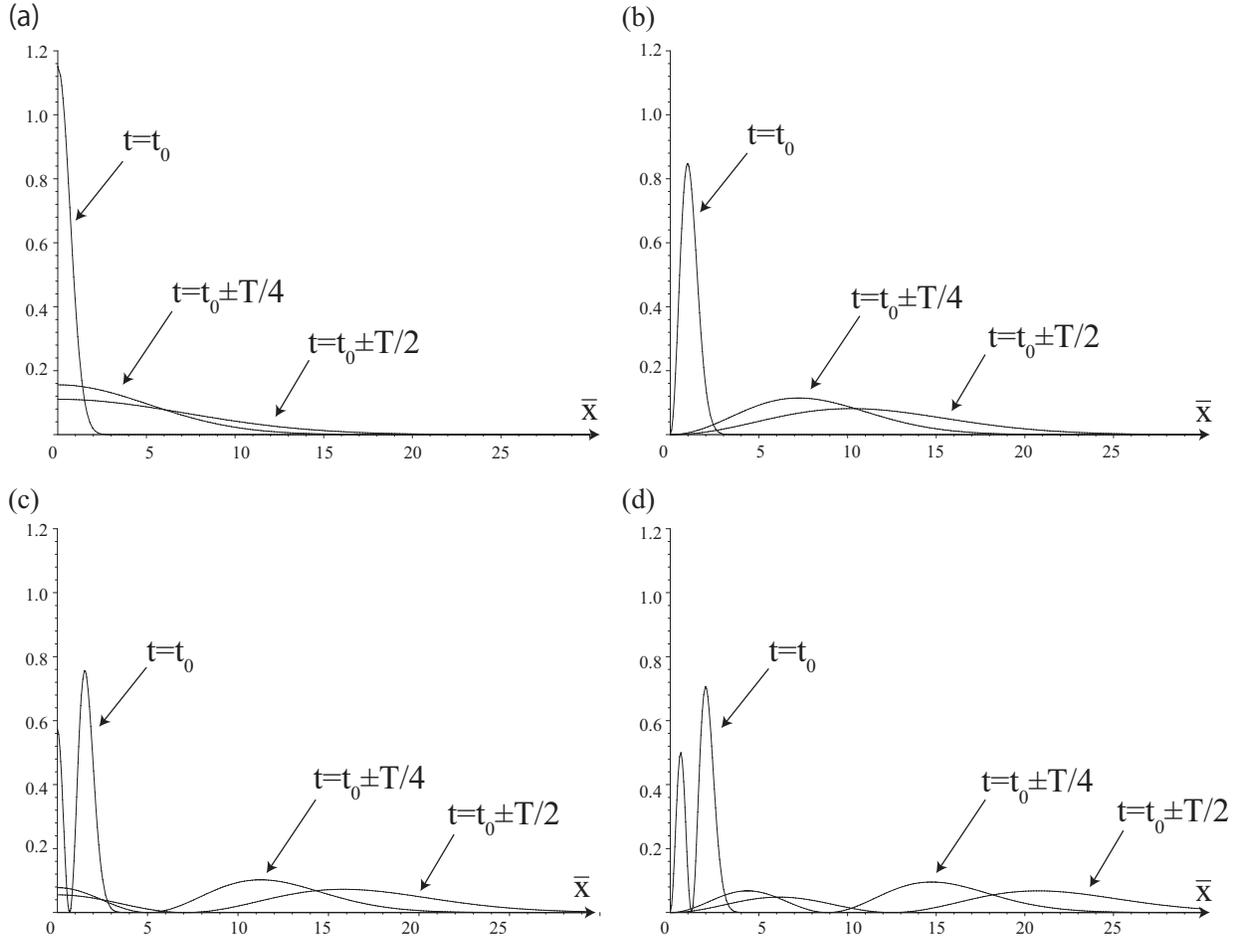}
\caption{\label{fig-wave1(all)} The evolution of the normalized squared norm $|\Psi_{{\rm I}(N)}({\bar x},t)|^2$ with $\alpha_0={\bar\beta}_0=1$ for (a) $N=0$, (b) $N=1$, (c) $N=2$, and (d) $N=3$.
$t_0$ is the time when $|\Psi_{{\rm I}(N)}({\bar x},t)|^2$ has the maximum value in the domain, and $T$ is the period of oscillation.
}
\end{center}
\end{figure*}

The solution (\ref{sol-first}) is a function in the Hilbert space ${\cal L}^2([0,\infty))$ with the Dirichlet or Neumann boundary condition at the origin and nonsingular for $t\in (-\infty,\infty)$.
Hence the classical initial singularity at $r=0$ (hence at $x={\bar x}=0$) is avoided in the quantum system.
The oscillatory evolution of the normalized squared norm $|\Psi_{{\rm I}(N)}({\bar x},t)|^2$ is exhibited in Fig.~\ref{fig-wave1(all)} for several values of the quantum number $N$.

We now examine the time evolution of the expectation value of the Euclidean volume constructed from the areal radius $r(=a)$, given by $V_{\rm E}=V_{n-2}^{(0)}x^2/(n-1)$.
Classically, along the orbit of the wormhole throat in the maximally extended spacetime, $V_{\rm E}(t)$ starts from zero at the white-hole singularity at $x=0$, reaches the maximum value at the bifurcation $(n-2)$-surface corresponding to the event horizon, and then turns to decrease toward $x=0$ at the black-hole singularity.

\begin{figure}[htbp]
\begin{center}
\includegraphics[width=0.3\linewidth]{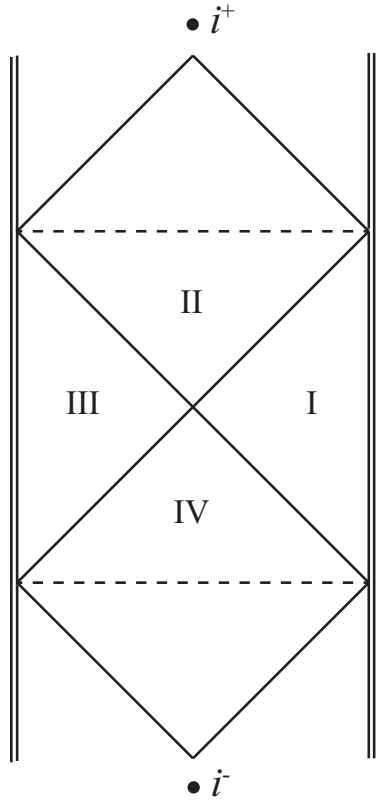}
\caption{\label{SingleBH-bounce} A portion of the Penrose diagram for the maximally extended spacetime of the regularized toroidal black hole. A double line corresponds to AdS infinity, and $i^+$ and $i^-$ are future and past timelike infinities, respectively. Dashed lines represent big bounces which replace the spacelike curvature singularities in classical general relativity.}
\end{center}
\end{figure}

The volume expectation value is given by 
\begin{align}
\langle V_{\rm E}\rangle_N(t)=&\frac{(2N+1)(n-1)\hbar \kappa_n^2}{8(n-2)\beta(t)^2} \nonumber \\
=&V_{\rm p}\frac{(2N+1)(n-1)l}{4(n-2)V_{n-2}^{(0)}\ell_{\rm p}}\frac{\Omega}{\omega}\biggl(1+\sqrt{1-\frac{\omega^2}{\Omega^2}}\sin (2\omega t+\theta_0)\biggl),\label{V_E-0}
\end{align}
where $\Omega$ is defined by Eq.~(\ref{Omega-def}) and we use the following expression:
\begin{align}
\frac{1}{\beta(t)^2}=&\frac{\Omega}{\omega^2}\biggl(1+\sqrt{1-\frac{\omega^2}{\Omega^2}}\sin (2\omega t+\theta_0)\biggl),\label{beta-inverse} \\
\theta_0:=&\arctan\biggl\{\frac{1}{2\alpha_0}\biggl(1-\frac{{\bar\beta}_0^2\Omega}{\omega^2}\biggl)\biggl\}.
\end{align}
(See Appendix~\ref{app:area} for derivation of $\langle V_{\rm E}\rangle_N$.)
This shows that $\langle V_{\rm E}\rangle_N$ is positive definite and oscillating, and hence the classical spacelike singularity at $r=0$ is resolved quantum mechanically and replaced by a big-bounce.
The classical description of the regularized black hole is presented in Fig.~\ref{SingleBH-bounce}.

Note that the combination of the parameters $\Omega$ defined by Eq.~(\ref{Omega-def}) provides a measure of the difference between the above states and the corresponding stationary state with the same $N$.
The expression
\begin{align}
\frac{\Omega}{\omega}=\frac{{\bar \beta}_0^4+4\alpha_0^2\omega^2+\omega^2}{2{\bar \beta}_0^2\omega}
\end{align}
shows that $\Omega\ge \omega$ holds, with equality holding only in the stationary case ($\alpha_0=0$ and ${\bar\beta}_0^2=\omega$).
On the other hand, the high-amplitude limit $\Omega/\omega\to \infty$ is realized in two independent limits: ${\bar\beta}_0^2/\omega\to\infty$ or $\alpha_0^2\omega/{\bar \beta}_0^2\to \infty$.
In the next subsection, we will see that this high-amplitude limit $\Omega/\omega\to \infty$ corresponds to infinite mass spacing, and therefore the limit is unphysical. 
The limit $\Omega/\omega\to \infty$ should therefore be interpreted as $\Omega\gg \omega$, as done in the following.

In terms of the Planck length $\ell_{\rm p}$ and the Planck volume $V_{\rm p}$, the maximum and minimum values of $\langle V_{\rm E}\rangle_N(t)$ are given by 
\begin{align}
\max_{t\in\mathbb{R}}\langle V_{\rm E}\rangle_N(t)=&\frac{(2N+1)(n-1)l}{4(n-2)V_{n-2}^{(0)}\ell_{\rm p}}\frac{\Omega}{\omega}\biggl(1+\sqrt{1-\frac{\omega^2}{\Omega^2}}\biggl)V_{\rm p},\\
\min_{t\in\mathbb{R}}\langle V_{\rm E}\rangle_N(t)=&\frac{(2N+1)(n-1)l}{4(n-2)V_{n-2}^{(0)}\ell_{\rm p}}\frac{\Omega}{\omega}\biggl(1-\sqrt{1-\frac{\omega^2}{\Omega^2}}\biggl)V_{\rm p},
\end{align}
where we use $\omega:=(n-1)/2l$.
In the stationary case ($\Omega=\omega$), $\langle V_{\rm E}\rangle_N$ is constant:
\begin{align}
\lim_{\Omega/\omega\to 1}\langle V_{\rm E}\rangle_N(t)=\frac{(2N+1)(n-1)l}{4(n-2)V_{n-2}^{(0)}\ell_{\rm p}}V_{\rm p}.
\end{align}
In the high-amplitude limit $\Omega/\omega\to \infty$, on the other hand, we obtain
\begin{align}
\lim_{\Omega/\omega\to \infty}\max_{t\in\mathbb{R}}\langle V_{\rm E}\rangle_N(t)\simeq& \frac{(2N+1)(n-1)l}{2(n-2)V_{n-2}^{(0)}\ell_{\rm p}}\frac{\Omega}{\omega}\to \infty,\\
\lim_{\Omega/\omega\to \infty}\min_{t\in\mathbb{R}}\langle V_{\rm E}\rangle_N(t)\simeq& \frac{(2N+1)(n-1)l}{8(n-2)V_{n-2}^{(0)}\ell_{\rm p}}\frac{\omega}{\Omega}\to 0.
\end{align}

\subsection{Mass of the  quantum black hole}
The physical meaning of the parameter $\Omega$ is made explicit by the following expression for the expectation value of the mass of the black hole:
\begin{align}
\langle M\rangle_N:=&\frac{\langle \Psi_{{\rm I}(N)}|{\hat H}\Psi_{{\rm I}(N)}\rangle}{\langle \Psi_{{\rm I}(N)}|\Psi_{{\rm I}(N)}\rangle}=\biggl(N+\frac12\biggl)\hbar \Omega. \label{mass-spectrum-dynamical}
\end{align}
(See Appendix.~\ref{app:mass} for derivation.)
Equation~(\ref{mass-spectrum-dynamical}) shows that $\Omega$, defined by Eq.~(\ref{Omega-def}), defines the mass step (i.e. spacing) between the neighboring states.
Since we have $\Omega\ge \omega$, with equality holding only for the stationary state, the mass step in the non-stationary state is larger than in the stationary state.

The expression (\ref{mass-spectrum-dynamical}) also confirms that the strict high-amplitude limit $\Omega/\omega\to \infty$ is unphysical.
However, we will still consider the situation $\Omega\gg \omega$ corresponding to a large black hole.
Although both of the limits $\Omega/\omega\to \infty$ and $N\to \infty$ correspond to the large-mass limit of a black hole, they are independent.

The mass spacing $\Delta\langle M\rangle_N$ is given by 
\begin{align}
\Delta\langle M\rangle_N:=\langle M\rangle_{N+1}-\langle M\rangle_N=\hbar \Omega
\end{align}
and hence $N\to\infty$ is a semi-classical limit in the sense that the relative mass spacing reduces to zero: 
\begin{align}
\lim_{N\to\infty}\frac{\Delta\langle M\rangle_N}{\langle M\rangle_N}=&\lim_{N\to\infty}\biggl(N+\frac12\biggl)^{-1}=0.
\end{align}
On the other hand, the mass fluctuation $\delta \langle M\rangle_N:=\sqrt{\langle M^2\rangle_N-\langle M\rangle_N^2}$ satisfies  
\begin{align}
\frac{\delta \langle M\rangle_N}{\langle M\rangle_N}=\sqrt{\frac{2(N^2+N+1)}{(2N+1)^2}\biggl(1-\frac{\omega^2}{\Omega^2}\biggl)}. \label{mass-uncertainty-dynamical}
\end{align}
(See Appendix~\ref{app:mass} for the derivation of $\delta \langle M\rangle_N$.)
Clearly, there is no mass uncertainty in the stationary case ($\Omega=\omega$).
In the non-stationary case, by contrast, the mass uncertainty is non-zero.
In the two independent large-mass limits $\Omega/\omega\to \infty$ and $N\to 0$, we obtain
\begin{align}
\lim_{\Omega/\omega\to \infty}\frac{\delta \langle M\rangle_N}{\langle M\rangle_N}=&\sqrt{\frac{2(N^2+N+1)}{(2N+1)^2}}\biggl(>\frac{1}{\sqrt{2}}\biggl),\\
\lim_{N\to\infty}\frac{\delta \langle M\rangle_N}{\langle M\rangle_N}=&\sqrt{\frac12\biggl(1-\frac{\omega^2}{\Omega^2}\biggl)}\biggl(<\frac{1}{\sqrt{2}}\biggl).
\end{align}
The relative mass fluctuations therefore are non-negligible even for large-mass quantum black holes with
$\Omega \gg \omega$.
On the other hand, for large black holes for which $N\to \infty$, the mass fluctuations can be small if the state is very close to the stationary state.

If  $\delta \langle M\rangle_N/\hbar\Omega\ge 1$, the mass fluctuation is greater than the spacing of the mass expectation value between states. 
Its physical implications are still not clear.
From the expression
\begin{align}
\frac{\delta \langle M\rangle_N}{\hbar\Omega}=\sqrt{\frac{N^2+N+1}{2}\biggl(1-\frac{\omega^2}{\Omega^2}\biggl)},
\end{align}
we see that $\delta \langle M\rangle_N/\hbar\Omega<1$ is satisfied for $N=0$, independent $\Omega$.
Also, the above expression shows that the condition $\delta \langle M\rangle_N/\hbar\Omega \ge 1$ is satisfied for $N\ge N_{\rm c}$, where the critical value $N_{\rm c}$ is a positive solution of the following algebraic equation:
\begin{align}
N_{\rm c}^2+N_{\rm c}+1=2\biggl(1-\frac{\omega^2}{\Omega^2}\biggl)^{-1}.
\end{align}
If $\Omega$ is close to $\omega$, then $\delta \langle M\rangle_N/\hbar\Omega<1$ is satisfied even for large $N$.
However, if the state is non-stationary, then $\delta \langle M\rangle_N/\hbar\Omega>1$ is satisfied only for sufficiently large $N$.

\subsection{Euclidean volume of the quantum black hole}
\label{app:volumeBH}

From Eqs.~(\ref{VC-def}) and (\ref{mass-spectrum-dynamical}), the Euclidean volume of the classical horizon $V_{\rm C}$ is given by 
\begin{align}
V_{\rm C}=&V_{\rm p}\frac{l^2(2N+1)\hbar\Omega}{(n-2)V_{n-2}^{(0)}l_{\rm p}^2m_{\rm p}}. \label{Vh}
\end{align}
By Eqs.~(\ref{deltaVC-def}) and (\ref{mass-uncertainty-dynamical}), the uncertainty $\delta V_{\rm C}$ satisfies 
\begin{align}
\frac{\delta V_{\rm C}}{V_{\rm C}}=\sqrt{\frac{2(N^2+N+1)}{(2N+1)^2}\biggl(1-\frac{\omega^2}{\Omega^2}\biggl)}. \label{deltaVh}
\end{align}
On the other hand, the Euclidean volume of the quantum horizon $V_{\rm Q}$ and its uncertainty $\delta V_{\rm Q}$ are defined by $V_{\rm Q}:=\langle V_{\rm E}\rangle_N(t_{\rm max})$ and $\delta V_{\rm Q}:=\delta \langle V_{\rm E}\rangle_N(t_{\rm max})$, respectively, where $t_{\rm max}$ is the time when $\langle V_{\rm E}\rangle_N$ has the maximum value.
Equation~(\ref{V_E-0}) can be written as
\begin{align}
\langle V_{\rm E}\rangle_N(t)=&V_{\rm p}\frac{l^2(2N+1)\hbar\Omega}{2(n-2)V_{n-2}^{(0)}\ell_{\rm p}^2m_{\rm p}} \biggl(1+\sqrt{1-\frac{\omega^2}{\Omega^2}}\sin (2\omega t+\theta_0)\biggl)
\end{align}
and its uncertainty is given by 
\begin{align}
\delta \langle V_{\rm E}\rangle_N(t)=&\sqrt{\frac{2(N^2+N+1)}{(2N+1)^2}}\langle V_{\rm E}\rangle(t) \label{deltaV}
\end{align}
(See in Appendix~\ref{app:area} for derivation.)
These expressions show
\begin{align}
V_{\rm Q}=&\frac12V_{\rm C}\biggl(1+\sqrt{1-\frac{\omega^2}{\Omega^2}}\biggl), \label{VQ} \\
\delta V_{\rm Q}=&V_{\rm Q}\sqrt{\frac{2(N^2+N+1)}{(2N+1)^2}} \nonumber \\
=&\frac12V_{\rm C}\sqrt{\frac{2(N^2+N+1)}{(2N+1)^2}}\biggl(1+\sqrt{1-\frac{\omega^2}{\Omega^2}}\biggl). \label{deltaVQ}
\end{align}

\begin{figure}[htbp]
\begin{center}
\includegraphics[width=0.7\linewidth]{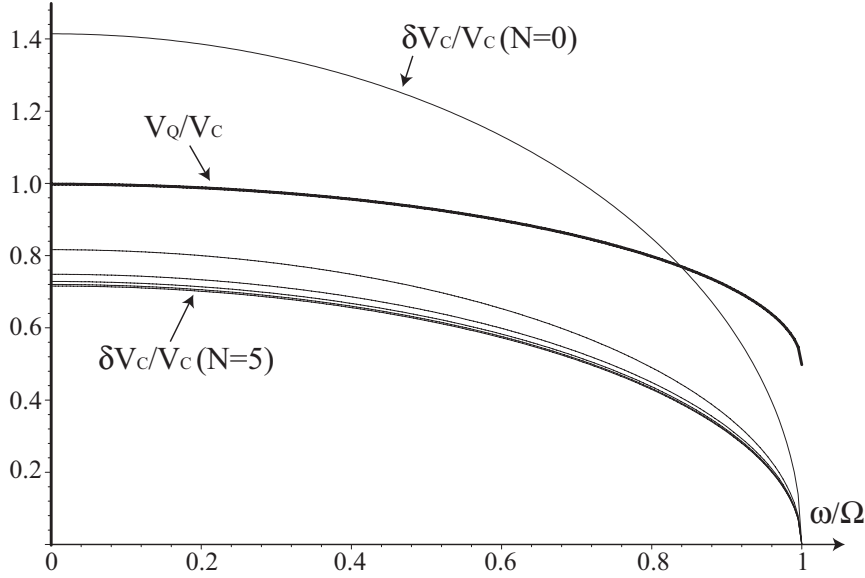}
\caption{\label{VQVC-deltaVC} Dependence of $V_{\rm Q}/V_{\rm C}$ (thick curve) and $\delta V_{\rm C}/V_{\rm C}$ for $N=0,1,\cdots,5$ (thin curves) on $\omega/\Omega$.
}
\end{center}
\end{figure}

Equations~(\ref{deltaVh}), (\ref{VQ}), and (\ref{deltaVQ}) imply the following relations:
\begin{align}
\frac{V_{\rm Q}}{V_{\rm C}}=&\frac12\biggl(1+\sqrt{1-\frac{\omega^2}{\Omega^2}}\biggl),\\
\frac{\delta V_{\rm C}}{V_{\rm C}}=&\frac{\delta V_{\rm Q}}{V_{\rm Q}}\sqrt{1-\frac{\omega^2}{\Omega^2}}, \label{V-relations}
\end{align}
which are independent of $N$.
$\delta V_{\rm C}/V_{\rm C}$ and $V_{\rm Q}/V_{\rm C}$ are drawn in Fig.~\ref{VQVC-deltaVC} as functions of $\omega/\Omega$ for $N=0,1,\cdots,5$.
We have $1/2\le V_{\rm Q}/V_{\rm C}<1$ and hence the quantum horizon is located inside the classical horizon.
The Euclidean volume of the quantum black hole is half that of the classical black hole in the stationary state ($\omega/\Omega=1$).
As $\omega/\Omega$ decreases, the ratio $V_{\rm Q}/V_{\rm C}$ monotonically increases and $V_{\rm Q}/V_{\rm C}=\delta V_{\rm Q}/\delta V_{\rm C}=1$ is satisfied, namely the quantum horizon coincides with the classical horizon, in the high-amplitude limit $\omega/\Omega\to 0$ (and hence for large black holes with $\Omega\gg \omega$).

Equation~(\ref{deltaVQ}) shows that $\delta V_{\rm Q}/V_{\rm Q}$ is independent of $\omega/\Omega$ and is a monotonically decreasing function of $N$ satisfying
\begin{align}
\frac{1}{\sqrt{2}}<\frac{\delta V_{\rm Q}}{V_{\rm Q}} \le \sqrt{2},
\end{align}
where the lower and upper bounds correspond to $N\to\infty$ and $N=0$, respectively.
Hence the relative volume fluctuation of the quantum horizon is not negligible even for large black holes described both by $N\to \infty$ and by $\Omega\gg\omega$.
This implies that the quantum black hole described by wave function I does not have a semi-classical limit.

On the other hand, Eq.~(\ref{deltaVh}) shows that $\delta V_{\rm C}/V_{\rm C}$ depends on $\omega/\Omega$ and is also a monotonically decreasing function of $N$ satisfying
\begin{align}
\sqrt{\frac{1}{2}\biggl(1-\frac{\omega^2}{\Omega^2}\biggl)}<\frac{\delta V_{\rm C}}{V_{\rm C}} \le \sqrt{2\biggl(1-\frac{\omega^2}{\Omega^2}\biggl)}.
\end{align}
As $\omega/\Omega$ becomes smaller, the properties of the classical horizon are more similar to the quantum horizon.
In contrast, the fluctuation of the classical horizon gets smaller and finally becomes zero in the stationary limit $\omega/\Omega\to 1$.

In closing this section, we present physical quantities in terms of the surface area for comparison.
By Eqs.~(\ref{def-Ac}), (\ref{beta-inverse}), and (\ref{mass-spectrum-dynamical}) and the results in Appendix~\ref{app:area}, we obtain
\begin{align}
\frac{A_{\rm Q}}{A_{\rm C}}=&\frac{\int_{0}^{\infty} y^{2(n-2)/(n-1)} e^{-y^2}H_N(y)^2 \D y}{2^{N-1}N!\sqrt{\pi}}\biggl\{\frac{1}{2N+1}\biggl(1+\sqrt{1-\frac{\omega^2}{\Omega^2}}\biggl)\biggl\}^{(n-2)/(n-1)}, \label{A-relations1}\\
\frac{\delta A_{\rm C}}{A_{\rm C}}=&\biggl\{\frac{2(N^2+N+1)}{(2N+1)^2}\biggl(1-\frac{\omega^2}{\Omega^2}\biggl)\biggl\}^{(n-2)/2(n-1)},\label{A-relations2}\\
\frac{\delta A_{\rm Q}}{A_{\rm Q}}=&\biggl(\frac{2^{N-1}N!\sqrt{\pi}\int_{0}^{\infty} y^{4(n-2)/(n-1)} e^{-y^2}H_N(y)^2 \D y}{(\int_{0}^{\infty} y^{2(n-2)/(n-1)} e^{-y^2}H_N(y)^2 \D y)^2}-1\biggl)^{1/2}. \label{A-relations3}
\end{align}
\begin{table}[htb]
\begin{center}
  \begin{tabular}{|l|c|c|c|c|c|c|} \hline
     & $A_{\rm Q}/A_{\rm C}|_{\omega/\Omega=1}$ & $A_{\rm Q}/A_{\rm C}|_{\omega/\Omega=0}$  & $\delta A_{\rm C}/A_{\rm C}|_{\omega/\Omega=1}$ & $\delta A_{\rm C}/A_{\rm C}|_{\omega/\Omega=0}$ & $\delta A_{\rm Q}/A_{\rm Q}$  \\ \hline \hline
    $n=3$ & 0.6366 & 0.9003  & 0 & 0.8409 & 0.4834   \\ 
    $n=4$ & 0.5798 & 0.9204  & 0 & 0.7937 & 0.5709  \\ 
    $n=5$ & 0.5564 & 0.9358  & 0 & 0.7711 & 0.6090   \\ 
    $n=6$ & 0.5436 & 0.9465  & 0 & 0.7579 & 0.6304  \\ 
    $n=7$ & 0.5356 & 0.9543  & 0 & 0.7492 & 0.6441   \\ 
    $n=8$ & 0.5301 & 0.9602  & 0 & 0.7430 & 0.6536  \\ 
    $n=9$ & 0.5260 & 0.9647  & 0 & 0.7384 & 0.6607  \\ 
    $n=10$ & 0.5229 & 0.9683  & 0 & 0.7349 & 0.666  \\ 
    $n\to\infty$ & 0.5 & 1  & 0 & 0.7071 & 0.7071  \\ 
 \hline
  \end{tabular}
  \caption{Values of $A_{\rm Q}/A_{\rm C}$, $\delta A_{\rm C}/A_{\rm C}$, and $\delta A_{\rm Q}/A_{\rm Q}$ for $\Psi=\Psi_{{\rm I}(N)}$ with $N=100$ are shown for $n=3,4,\cdots,10$ and $n\to\infty$. The values of $\omega/\Omega$ in $A_{\rm Q}/A_{\rm C}$ and $\delta A_{\rm C}/A_{\rm C}$ are $\omega/\Omega=1$ and $0$. In the limit $n\to\infty$, the values are identical to the ones in terms of the Euclidean volume with $N=100$, shown in Table~\ref{Table:largeN-V}.}
\label{Table:largeN}
  \begin{tabular}{|l|c|c|c|c|c|c|} \hline
     & $V_{\rm Q}/V_{\rm C}|_{\omega/\Omega=1}$ & $V_{\rm Q}/V_{\rm C}|_{\omega/\Omega=0}$  & $\delta V_{\rm C}/V_{\rm C}|_{\omega/\Omega=1}$ & $\delta V_{\rm C}/V_{\rm C}|_{\omega/\Omega=0}$ & $\delta V_{\rm Q}/V_{\rm Q}$  \\ \hline \hline
    $N=0$ & 0.5 & 1  & 0 & 1.414 & 1.414   \\ 
    $N=1$ & 0.5 & 1  & 0 & 0.8165 & 0.8165  \\ 
    $N=2$ & 0.5 & 1  & 0 & 0.7483 & 0.7483   \\ 
    $N=3$ & 0.5 & 1  & 0 & 0.7284 & 0.7284  \\ 
    $N=4$ & 0.5 & 1  & 0 & 0.7201 & 0.7201   \\ 
    $N=5$ & 0.5 & 1  & 0 & 0.7158 & 0.7158   \\ 
    $N=10$ & 0.5 & 1  & 0 & 0.7095 & 0.7095  \\ 
    $N=100$ & 0.5 & 1  & 0 & 0.7071 & 0.7071  \\ 
    $N\to \infty$ & 0.5 & 1  & 0 & 0.7071 & 0.7071  \\ 
 \hline
  \end{tabular}
  \caption{Values of $V_{\rm Q}/V_{\rm C}$, $\delta V_{\rm C}/V_{\rm C}$, and $\delta V_{\rm Q}/V_{\rm Q}$ are shown for $\Psi=\Psi_{{\rm I}(N)}$ for some values of $N$. The values of $\omega/\Omega$ in $V_{\rm Q}/V_{\rm C}$ and $\delta V_{\rm C}/V_{\rm C}$ are $\omega/\Omega=1$ and $0$.}
\label{Table:largeN-V}
\end{center}
\end{table}
The values of $A_{\rm Q}/A_{\rm C}$, $\delta A_{\rm C}/A_{\rm C}$, and $\delta A_{\rm Q}/A_{\rm Q}$ in the large-$N$ (actually $N=100$) case are shown in Table~\ref{Table:largeN}, where $\omega/\Omega$ is set to be $1$ and $0$ for $A_{\rm Q}/A_{\rm C}$ and $\delta A_{\rm C}/A_{\rm C}$.
For comparison, we present the same quantities using the Euclidean volume instead of area in Table~\ref{Table:largeN-V}.

\section{Quantum toroidal black hole: Wave function II}
\label{sec:exact2}

In this section, we will study quantum black holes described by the wave function $\Psi=\Psi^{(\pm)}_{{\rm II}(N)}$, which is given by Eq.~(\ref{SolHalfLine}) with $\alpha_0=0$ and ${\bar \beta}_0=\sqrt{\omega}$ and corresponds to generalized coherent states on the half line.
In the present section, the upper and lower signs in the expression of the expectation value $\langle X\rangle_{N}$ for a quantity $X$ are for $\Psi=\Psi^{(+)}_{{\rm II}(N)}$ and $\Psi=\Psi^{(-)}_{{\rm II}(N)}$, respectively.

\subsection{Dynamical singularity resolution}
Let us compute the expectation value of the Euclidean volume with the areal radius $r(=a)$:
\begin{align}
\langle V_{\rm E}\rangle_N(t):=&\frac{V_{n-2}^{(0)}}{n-1}\langle x^{2}\rangle_N =V_{\rm p}\frac{(n-1)l\omega\langle {\bar x}^2\rangle_N}{2(n-2)V_{n-2}^{(0)}\ell_{\rm p}}. \label{<V>E-wave2}
\end{align}
By construction, $\langle V_{\rm E}\rangle_N(t)$ is positive definite and hence the classical singularity at $r=0$ (hence ${\bar x}=0$) is resolved quantum mechanically.

The best friendly forms of the first six of the quantity $\omega\langle {\bar x}^2\rangle_N$ are
\begin{align}
\omega\langle {\bar x}^2\rangle_{0}=&\frac{2\varepsilon^2+1}{2}\mp \frac{\zeta}{e^\zeta\pm 1} \nonumber \\
\omega\langle {\bar x}^2\rangle_1=&\frac{2\varepsilon^2+3}{2}\mp \frac{4\varepsilon^2+\zeta(2\zeta-3)}{e^\zeta\pm(2\zeta-1)},\label{<X2>-II1}\\
\omega\langle {\bar x}^2\rangle_2=&\frac{2\varepsilon^2+5}{2}\mp \frac{4(2\zeta-3)\varepsilon^2+\zeta(2\zeta^2-8\zeta+7)}{e^\zeta\pm(2\zeta^2-4\zeta+1)},\\
\omega\langle {\bar x}^2\rangle_3=&\frac{2\varepsilon^2+7}{2} \mp \frac{24(\zeta-1)(\zeta-3)\varepsilon^2+\zeta(4\zeta^3-30\zeta^2+66\zeta-39)}{3e^\zeta\pm(4\zeta^3-18\zeta^2+18\zeta-3)},\\
\omega\langle {\bar x}^2\rangle_4=&\frac{2\varepsilon^2+9}{2}\mp \frac{8(2\zeta^3-15\zeta^2+30\zeta-15)\varepsilon^2+\zeta(2\zeta^4-24\zeta^3+96\zeta^2-144\zeta+63)}{3e^\zeta\pm(2\zeta^4-16\zeta^3+36\zeta^2-24\zeta+3)},\\
\omega\langle {\bar x}^2\rangle_5=&\frac{2\varepsilon^2+11}{2}  \nonumber \\
&\mp \frac{20(2\zeta^4-24\zeta^3+90\zeta^2-120\zeta+45)\varepsilon^2+\zeta(4\zeta^5-70\zeta^4+440\zeta^3-1200\zeta^2+1350\zeta-465)}{15e^\zeta\pm(4\zeta^5-50\zeta^4+200\zeta^3-300\zeta^2+150\zeta-15)}.\label{<X2>-II2}
\end{align}
It is seen that the time dependence of $\langle V_{\rm E}\rangle_N$ is determined by $\varepsilon(t)^2$.
Because $\varepsilon(t)$ itself is oscillating within the range $-\sqrt{\zeta}\le \varepsilon\le\sqrt{\zeta}$ as shown in Eq.~(\ref{eq:SPepsilon}), the dynamics of $\langle V_{\rm E}\rangle_N$ is a simple oscillation.
The above expressions imply
\begin{align}
\lim_{\zeta\to\infty}\omega\langle {\bar x}^2\rangle_N\simeq &\varepsilon^2+\biggl(N+\frac12\biggl)+{\cal O}(\zeta^{N+1}e^{-\zeta}) \label{limit<x^2>}
\end{align}
and hence $\langle V_{\rm E}\rangle_N$ is almost equally spaced for large $\zeta$.

\subsection{Mass of quantum black hole}
The first six mass expectation values $\langle M\rangle_N$ are given by 
\begin{align}
\langle M\rangle_0=&\frac12\hbar\omega+\frac12\hbar\omega\zeta\frac{e^{\zeta}\mp1}{e^{\zeta}\pm1},\label{<M>0-II}\\
\langle M\rangle_1=&\frac32\hbar\omega+\frac12\hbar\omega\zeta\frac{e^\zeta\mp(2\zeta-1)}{e^{\zeta}\pm(2\zeta-1)},\\
\langle M\rangle_2=&\frac52\hbar\omega+\frac12\hbar\omega\zeta\frac{e^\zeta\mp(2\zeta^2-4\zeta+1)}{e^{\zeta}\pm(2\zeta^2-4\zeta+1)},\\
\langle M\rangle_3=&\frac72\hbar\omega+\frac12\hbar\omega\zeta\frac{3e^\zeta\mp(4\zeta^3-18\zeta^2+18\zeta-3)}{3e^{\zeta}\pm(4\zeta^3-18\zeta^2+18\zeta-3)},\\
\langle M\rangle_4=&\frac92\hbar\omega+\frac12\hbar\omega\zeta\frac{3e^\zeta\mp(2\zeta^4-16\zeta^3+36\zeta^2-24\zeta+3)}{3e^{\zeta}\pm(2\zeta^4-16\zeta^3+36\zeta^2-24\zeta+3)},\\
\langle M\rangle_5=&\frac{11}{2}\hbar\omega+\frac12\hbar\omega\zeta\frac{15e^\zeta\mp(4\zeta^5-50\zeta^4+200\zeta^3-300\zeta^2+150\zeta-15)}{15e^{\zeta}\pm(4\zeta^5-50\zeta^4+200\zeta^3-300\zeta^2+150\zeta-15)}.\label{<M>5-II}
\end{align}
\begin{figure*}[htbp]
\begin{center}
\includegraphics[width=0.65\linewidth]{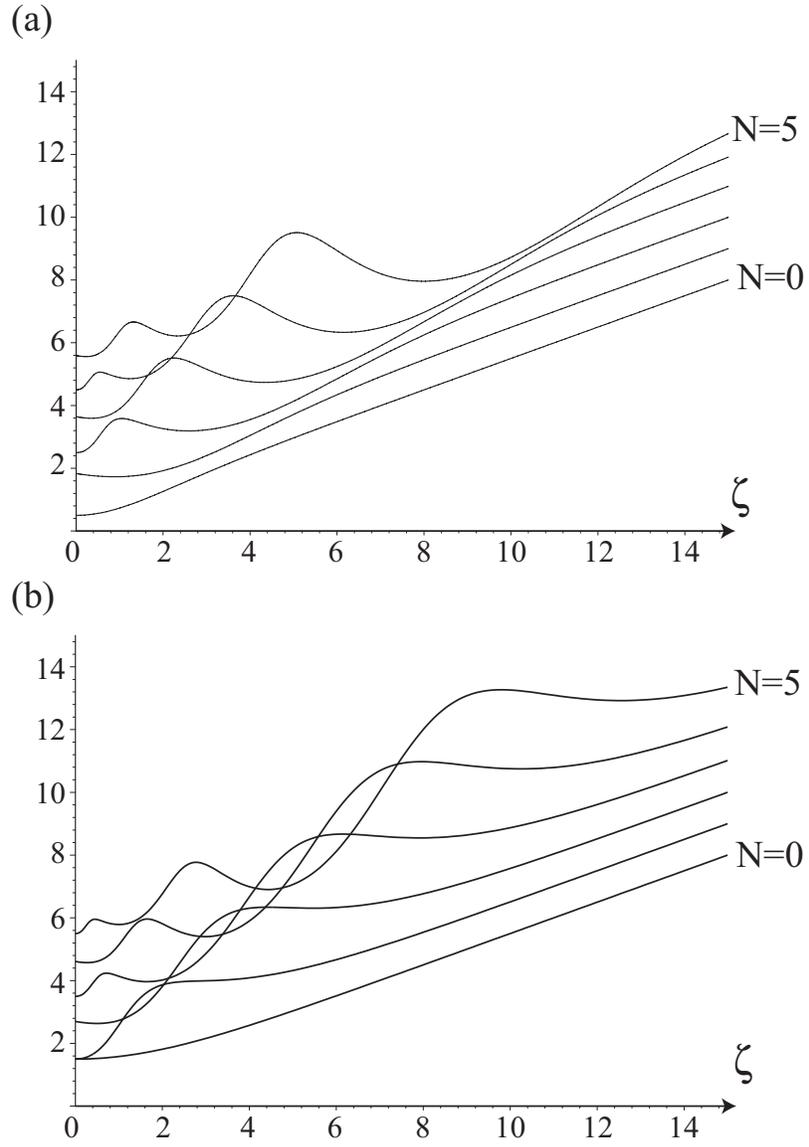}
\caption{\label{fig-Mcoherent} The first six mass expectation values $\langle M\rangle_N/\hbar\omega$ as a function of $\zeta$ for (a) $\Psi=\Psi^{(+)}_{{\rm II}(N)}$ and (b) $\Psi=\Psi^{(-)}_{{\rm II}(N)}$.
}
\end{center}
\end{figure*}
The forms of $\langle M\rangle_N$ are drawn in Fig.~\ref{fig-Mcoherent}.
It shows that $\langle M\rangle_N$ is not a monotonic function of $\zeta$ in general and $\langle M\rangle_N$ is almost equally spaced for large $\zeta$, shown as
\begin{align}
\lim_{\zeta\to\infty}\langle M\rangle_N\simeq&\biggl(N+\frac12\biggl)\hbar\omega+\frac12\hbar\omega\zeta+{\cal O}(\zeta^{-1}). \label{<M>largezeta}
\end{align}
For small $\zeta$, on the other hand, $\langle M\rangle_{N+1}$ can be smaller than $\langle M\rangle_N$ in some ranges of $\zeta$.
It is intriguing that this type of level crossing occurs in time-dependent states of quantum black holes.
\begin{figure*}[htbp]
\begin{center}
\includegraphics[width=0.65\linewidth]{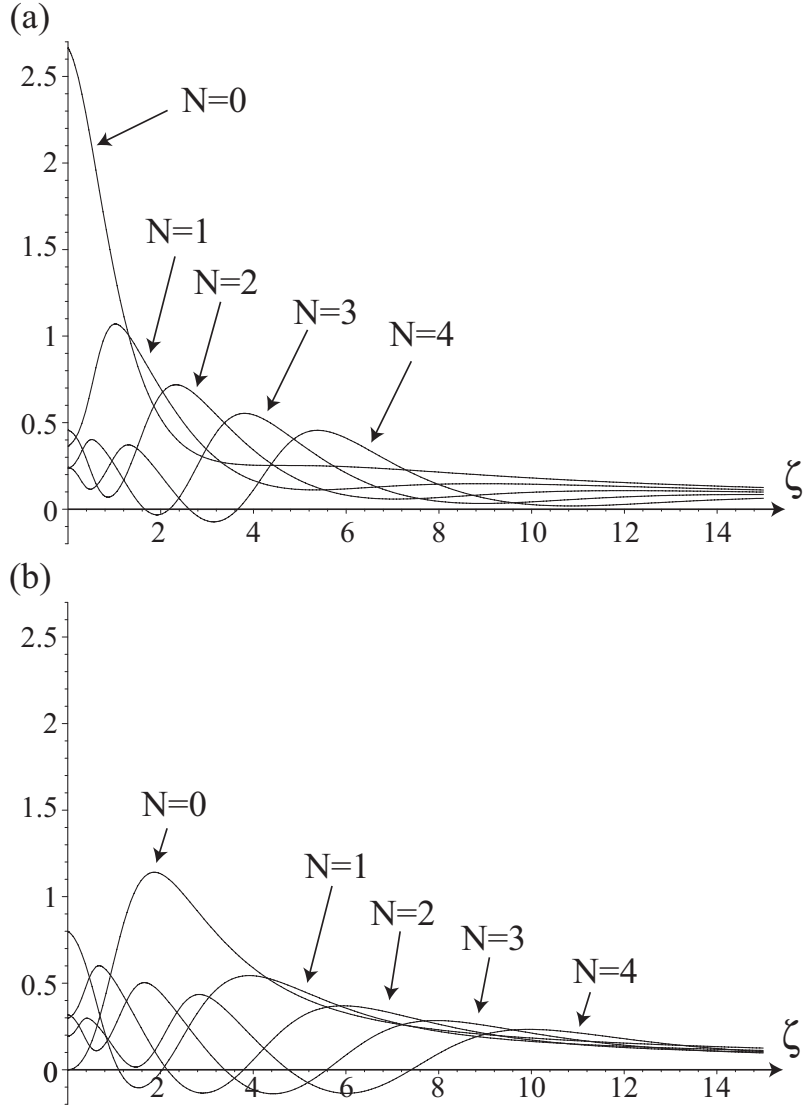}
\caption{\label{fig-Mspacing} The first five relative mass spacings $\Delta \langle M\rangle_N/\langle M\rangle_{N}$ as functions of $\zeta$ for (a) $\Psi=\Psi^{(+)}_{{\rm II}(N)}$ and (b) $\Psi=\Psi^{(-)}_{{\rm II}(N)}$.
}
\end{center}
\end{figure*}


It is noted that all of the above $\langle M\rangle_N$ have finite limits for $\zeta\to 0$, in spite of the fact that $\Psi^{(\pm)}_{II(N)}\equiv 0$ is realized in some cases.
The limits are 
\begin{align}
\lim_{\zeta\to 0}\langle M\rangle_0=&\frac12\hbar\omega,\quad \lim_{\zeta\to 0}\langle M\rangle_1=\frac{11}{6}\hbar\omega,\\
\lim_{\zeta\to 0}\langle M\rangle_2=&\frac52\hbar\omega,\quad \lim_{\zeta\to 0}\langle M\rangle_3=\frac{51}{14}\hbar\omega,\\
\lim_{\zeta\to 0}\langle M\rangle_4=&\frac92\hbar\omega,\quad \lim_{\zeta\to 0}\langle M\rangle_5=\frac{123}{22}\hbar\omega
\end{align}
for $\Psi=\Psi^{(+)}_{{\rm II}(N)}$ (where the states with $N=1,3,5$ are unphysical for $\zeta\to 0$), and 
\begin{align}
\lim_{\zeta\to 0}\langle M\rangle_0=&\frac32\hbar\omega,\quad \lim_{\zeta\to 0}\langle M\rangle_1=\frac32\hbar\omega, \\
\lim_{\zeta\to 0}\langle M\rangle_2=&\frac{27}{10}\hbar\omega,\quad \lim_{\zeta\to 0}\langle M\rangle_3=\frac72\hbar\omega,\\
\lim_{\zeta\to 0}\langle M\rangle_4=&\frac{83}{18}\hbar\omega,\quad \lim_{\zeta\to 0}\langle M\rangle_5=\frac{11}{2}\hbar\omega
\end{align}
for $\Psi=\Psi^{(-)}_{{\rm II}(N)}$ (where the states with $N=0,2,4$ are unphysical for $\zeta\to 0$).

The expressions (\ref{<M>0-II})--(\ref{<M>5-II}) and Eq.~(\ref{<M>largezeta}) imply that $N\to \infty$ and $\zeta\to \infty$ are independent large black-hole limits.
The relative mass spacing $\Delta \langle M\rangle_N/\langle M\rangle_{N}$ is drawn in Fig.~\ref{fig-Mspacing}, which shows that $\Delta \langle M\rangle_N/\langle M\rangle_{N}$ converges to zero for $N\to \infty$ and also for $\zeta\gg N$. 
Indeed, the expressions (\ref{<M>0-II})--(\ref{<M>5-II}) imply
\begin{align}
\lim_{\zeta\to\infty}\frac{\Delta \langle M\rangle_N}{\langle M\rangle_N} \propto \frac{1}{\zeta}\to0.
\end{align}
Here it is emphasized that the limit $\zeta\to\infty$ in the following argument does not mean infinite $\zeta$ but sufficiently large $\zeta$ satisfying $N/\zeta\ll 1$.
We will now examine the behaviour of the quantum fluctuation in the limit of $N\to \infty$ or $\zeta\to \infty$.

\begin{figure}[htbp]
\begin{center}
\includegraphics[width=0.7\linewidth]{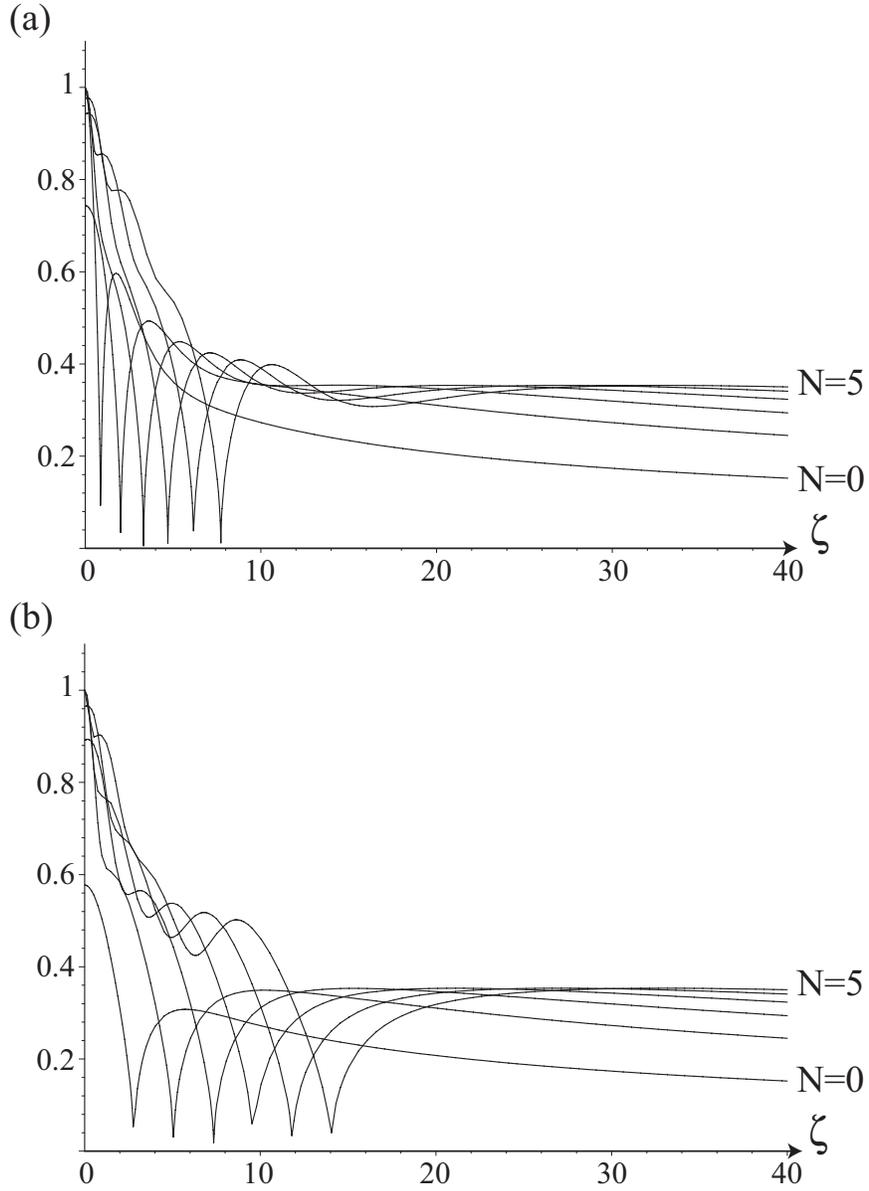}
\caption{\label{fig-deltaMcoherent} The first six relative mass fluctuations $\delta\langle M\rangle_N/\langle M\rangle_N$ as functions of $\zeta$ for (a) $\Psi=\Psi^{(+)}_{{\rm II}(N)}$ and (b) $\Psi=\Psi^{(-)}_{{\rm II}(N)}$. }
\end{center}
\end{figure}

The relative mass fluctuation is given by 
\begin{align}
\frac{\delta\langle M\rangle_N}{\langle M\rangle_N}=&\frac{\sqrt{|\langle M^2\rangle_N-\langle M\rangle_N^2|}}{\langle M\rangle_N} =\sqrt{\biggl|\frac{\int_{0}^\infty \Psi^*{\hat H}^2\Psi\D {\bar x}}{(\int_{0}^\infty\Psi^*{\hat H}\Psi\D {\bar x})^2}-1\biggl|},
\end{align}
where $\Psi=\Psi^{(\pm)}_{{\rm II}(N)}$.
The simplest forms of the first six are given by 
\begin{align}
\biggl(\frac{\delta\langle M\rangle_0}{\langle M\rangle_0}\biggl)^2=&\biggl|\frac{\zeta-1}{(\zeta+1)^2}-\frac{2\zeta h_0(\zeta)}{(\zeta+1)^2\{(\zeta+1)e^{\zeta}\mp (\zeta-1)\}^2}\biggl|,\\
\biggl(\frac{\delta\langle M\rangle_1}{\langle M\rangle_1}\biggl)^2=&\biggl|\frac{3(\zeta-3)}{(\zeta+3)^2}-\frac{2\zeta h_1(\zeta)}{(\zeta+3)^2\{(\zeta+3)e^{\zeta}\mp (\zeta-3)(2\zeta-1)\}^2}\biggl|,\\
\biggl(\frac{\delta\langle M\rangle_2}{\langle M\rangle_2}\biggl)^2=&\biggl|\frac{5(\zeta-5)}{(\zeta+5)^2}-\frac{2\zeta h_2(\zeta)}{(\zeta+5)^2\{(\zeta+5)e^{\zeta}\mp (\zeta-5)(2\zeta^2-4\zeta+1)\}^2}\biggl|,\\
\biggl(\frac{\delta\langle M\rangle_3}{\langle M\rangle_3}\biggl)^2=&\biggl|\frac{7(\zeta-7)}{(\zeta+7)^2}-\frac{2\zeta h_3(\zeta)}{(\zeta+7)^2\{3(\zeta+7)e^{\zeta}\mp (\zeta-7)(4\zeta^3-18\zeta^2+18\zeta-3)\}^2}\biggl|,\\
\biggl(\frac{\delta\langle M\rangle_4}{\langle M\rangle_4}\biggl)^2=&\biggl|\frac{9(\zeta-9)}{(\zeta+9)^2}-\frac{2\zeta h_4(\zeta)}{(\zeta+9)^2\{3(\zeta+9)e^{\zeta}\mp (\zeta-9)(2\zeta^4-16\zeta^3+36\zeta^2-24\zeta+3)\}^2}\biggl|,\\
\biggl(\frac{\delta\langle M\rangle_5}{\langle M\rangle_5}\biggl)^2=&\biggl|\frac{11(\zeta-11)}{(\zeta+11)^2} \nonumber \\
&-\frac{2\zeta h_5(\zeta)}{(\zeta+11)^2\{15(\zeta+11)e^{\zeta}\mp (\zeta-11)(4\zeta^5-50\zeta^4+200\zeta^3-300\zeta^2+150\zeta-15)\}^2}\biggl|,
\end{align}
where
\begin{align}
h_0:=&(\zeta^2+3)\mp (\zeta+1)(2\zeta^2+3\zeta-3)e^\zeta,\\
h_1:=&(2\zeta-1)(2\zeta^3-27\zeta^2+126\zeta-81)\mp (\zeta-1)(\zeta+3)(4\zeta^2+24\zeta-27)e^\zeta,\\
h_2:=&(2\zeta^2-4\zeta+1)(2\zeta^4-88\zeta^3+675\zeta^2-1200\zeta+375) \nonumber \\
&\mp (\zeta+5)(4\zeta^4+30\zeta^3-180\zeta^2+255\zeta-75) e^\zeta,\\
h_3:=&(4\zeta^3-18\zeta^2+18\zeta-3)(4\zeta^5-366\zeta^4+4338\zeta^3-14847\zeta^2+14994\zeta-3087) \nonumber \\
&\mp 3(\zeta+7)(8\zeta^5+72\zeta^4-858\zeta^3+2418\zeta^2-2205\zeta+441)e^\zeta,\\
h_4:=&(2\zeta^4-16\zeta^3+36\zeta^2-24\zeta+3) \nonumber \\
&\times(2\zeta^6-312\zeta^5+5322\zeta^4-29688\zeta^3+61479\zeta^2-42768\zeta+6561) \nonumber \\
&\mp 3(\zeta+9)(4\zeta^6+38\zeta^5-822\zeta^4+3972\zeta^3-7350\zeta^2+4833\zeta-729)e^\zeta,\\
h_5:=&(4\zeta^5-50\zeta^4+200\zeta^3-300\zeta^2+150\zeta-15) \nonumber \\
&\times(4\zeta^7-950\zeta^6+22092\zeta^5-182070\zeta^4+634350\zeta^3-927465\zeta^2+490050\zeta-59895) \nonumber \\
&\mp 15(\zeta+11)(8\zeta^7+72\zeta^6-2742\zeta^5+20670\zeta^4-64800\zeta^3+88320\zeta^2-45045\zeta+5445)e^\zeta.
\end{align}
The forms of $\delta\langle M\rangle_N/\langle M\rangle_N$ are drawn in Fig.~\ref{fig-deltaMcoherent}.
The second term in the above expressions of $\delta\langle M\rangle_N/\langle M\rangle_N$ rapidly converges to zero as $\zeta$ increases and therefore the fall-off rate of the fluctuation should be 
\begin{align}
\lim_{\zeta\to\infty}\frac{\delta\langle M\rangle_N}{\langle M\rangle_N}\simeq& \frac{(2N+1)\{\zeta-(2N+1)\}}{(\zeta+2N+1)^2}+{\cal O}(\zeta^Ne^{-\zeta}) \to 0. \label{delta<M>largezeta}
\end{align}
Hence, the relative mass fluctuation of the black hole in the non-stationary state becomes very small in a large-mass limit $\zeta\gg N$.
In contrast, as seen in Fig.~\ref{fig-deltaMcoherent}, the relative mass fluctuation is not negligible in a different large-mass limit $N\to\infty$.

\subsection{Euclidean volume of quantum black hole}
The Euclidean volume $\langle V_{\rm E}\rangle_N$ with the areal radius $r=a(t)$ is given by Eq.~(\ref{<V>E-wave2}), and $\omega\langle {\bar x}^2\rangle_N$ has the form of Eq.~(\ref{limit<x^2>}) for large $\zeta$.
In this subsection, we study the Euclidean volume of the classical and quantum horizons in the case of large $\zeta$, in which we can obtain their analytic expressions.

For large $\zeta$, the volume of the quantum horizon $V_{\rm Q}$ behaves as 
\begin{align}
\lim_{\zeta\to\infty} V_{\rm Q}=&\lim_{\zeta\to\infty} \max_{t\in\mathbb{R}}\langle V_{\rm E}\rangle_N(t) \nonumber \\
\simeq& \frac{(n-1)V_{\rm p}l(2\zeta+2N+1)}{4(n-2)V_{n-2}^{(0)}\ell_{\rm p}}\to \infty.\label{VQ2-large}
\end{align}
In order to see the volume fluctuation of the quantum horizon $\delta V_{\rm Q}/V_{\rm Q}$, we need to compute
\begin{align}
\frac{\delta\langle V_{\rm E}\rangle_N}{\langle V_{\rm E}\rangle_N}=&\frac{\sqrt{|\langle V_{\rm E}^2\rangle_N-\langle V_{\rm E}\rangle_N^2|}}{\langle V_{\rm E}\rangle_N} =\sqrt{\biggl|\frac{\int_{0}^\infty {\bar x}^4|\Psi|^2\D {\bar x}}{(\int_{0}^\infty {\bar x}^2|\Psi|^2\D {\bar x})^2}-1\biggl|},
\end{align}
where $\Psi=\Psi^{(\pm)}_{{\rm II}(N)}$.
Although it is difficult to obtain the expression for general $N$, the first six are written as 
\begin{align}
\biggl(\frac{\delta\langle V_{\rm E}\rangle_0}{\langle V_{\rm E}\rangle_0}\biggl)^2=&\biggl|\frac{2(1+4\varepsilon^2)}{(1+2\varepsilon^2)^2}+j_{0(\pm)}(\zeta,\varepsilon^2)\biggl|,\\
\biggl(\frac{\delta\langle V_{\rm E}\rangle_1}{\langle V_{\rm E}\rangle_1}\biggl)^2=&\biggl|\frac{2(3+12\varepsilon^2)}{(3+2\varepsilon^2)^2}+j_{1(\pm)}(\zeta,\varepsilon^2)\biggl|,\\
\biggl(\frac{\delta\langle V_{\rm E}\rangle_2}{\langle V_{\rm E}\rangle_2}\biggl)^2=&\biggl|\frac{2(7+20\varepsilon^2)}{(5+2\varepsilon^2)^2}+j_{2(\pm)}(\zeta,\varepsilon^2)\biggl|,\\
\biggl(\frac{\delta\langle V_{\rm E}\rangle_3}{\langle V_{\rm E}\rangle_3}\biggl)^2=&\biggl|\frac{2(13+28\varepsilon^2)}{(7+2\varepsilon^2)^2}+j_{3(\pm)}(\zeta,\varepsilon^2)\biggl|,\\
\biggl(\frac{\delta\langle V_{\rm E}\rangle_4}{\langle V_{\rm E}\rangle_4}\biggl)^2=&\biggl|\frac{2(21+36\varepsilon^2)}{(9+2\varepsilon^2)^2}+j_{4(\pm)}(\zeta,\varepsilon^2)\biggl|,\\
\biggl(\frac{\delta\langle V_{\rm E}\rangle_5}{\langle V_{\rm E}\rangle_5}\biggl)^2=&\biggl|\frac{2(31+44\varepsilon^2)}{(11+2\varepsilon^2)^2}+j_{4(\pm)}(\zeta,\varepsilon^2)\biggl|,
\end{align}
where the expressions of the functions $j_{N(\pm)}(\zeta,\varepsilon^2)$ are complicated but converge to zero for large $\zeta$ as 
\begin{align}
\lim_{\zeta\to \infty}j_{N(\pm)}\simeq &\frac{\zeta^{N+2}e^{-\zeta}}{\{(2N+1)+2\varepsilon^2\}^2}\to0.
\end{align}
Hence, the above expressions imply that the relative volume fluctuation should behave as
\begin{align}
\lim_{\zeta\to\infty}\biggl(\frac{\delta\langle V_{\rm E}\rangle_N}{\langle V_{\rm E}\rangle_N}\biggl)^2\simeq &\frac{2(N^2+N+1)+8(2N+1)\varepsilon^2}{\{(2N+1)+2\varepsilon^2\}^2}+{\cal O}(\zeta^{N+2}e^{-\zeta}).
\end{align}
The leading term is exactly the same as in the shape-preserving state on the whole line.
Thus, the relative volume fluctuation of the quantum horizon for large $\zeta$ should be given by 
\begin{align}
\lim_{\zeta\to\infty}\frac{\delta V_{\rm Q}}{V_{\rm Q}}\simeq &\frac{\sqrt{2(N^2+N+1)+8(2N+1)\zeta}}{(2N+1)+2\zeta} \to 0.
\end{align}

On the other hand, by Eqs.~(\ref{VC-def}) and (\ref{<M>largezeta}), we obtain the Euclidean volume of the classical horizon $V_{\rm C}$ for large $\zeta$ as 
\begin{align}
\lim_{\zeta\to\infty}V_{\rm C}\simeq&\frac{(n-1)(\zeta+2N+1)V_{\rm p}l}{2(n-2)V_{n-2}^{(0)}\ell_{\rm p}}\to \infty.\label{VC2-large}
\end{align}
Equations~(\ref{deltaVC-def}) and (\ref{delta<M>largezeta}) show
\begin{align}
\lim_{\zeta\to\infty}\frac{\delta V_{\rm C}}{V_{\rm C}}\simeq \frac{(2N+1)\{\zeta-(2N+1)\}}{(\zeta+2N+1)^2}\to 0.
\end{align}

Now we have shown that the relative volume fluctuations of the horizon are very small in a large-mass limit $\zeta\gg N$.
In addition, Eqs.~(\ref{VQ2-large}) and (\ref{VC2-large}) show that $V_{\rm Q}/V_{\rm C}$ is close to $1$, shown as
\begin{align}
\lim_{\zeta\to\infty} \frac{V_{\rm Q}}{V_{\rm C}}\simeq& \frac{2\zeta+2N+1}{2(\zeta+2N+1)}\simeq 1-\frac{N}{\zeta}.
\end{align}
These results show that classical configurations are realized in this limit.
\begin{figure*}[htbp]
\begin{center}
\includegraphics[width=1.0\linewidth]{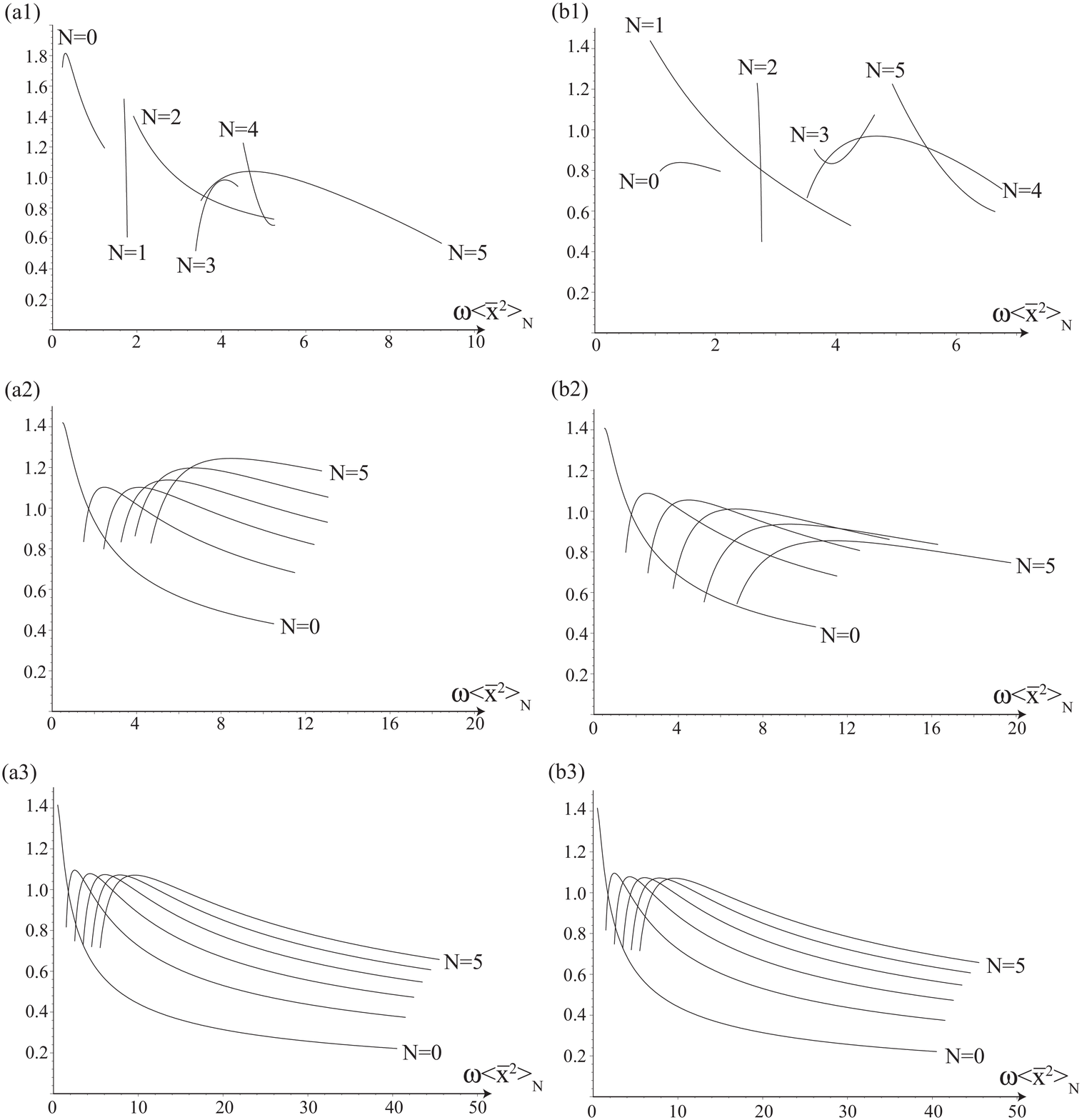}
\caption{\label{fig-deltaV(all)zeta=1-10-40} Values of $\delta \langle V_{\rm E}\rangle_N/\langle V_{\rm E}\rangle_N$ in the range of $\min_{t\in\mathbb{R}}\omega\langle {\bar x}^2\rangle_N\le \omega\langle {\bar x}^2\rangle_N\le \max_{t\in\mathbb{R}}\omega\langle {\bar x}^2\rangle_N$ for $N=0,\cdots 5$.
The left row corresponds to $\Psi=\Psi^{(+)}_{{\rm II}(N)}$, with (a1) $\zeta=1$, (a2) $\zeta=10$, and (a3) $\zeta=40$.
The right row corresponds to $\Psi=\Psi^{(-)}_{{\rm II}(N)}$, with (b1) $\zeta=1$, (b2) $\zeta=10$, and (b3) $\zeta=40$.
}
\end{center}
\end{figure*}

On the other hand, in another large-mass limit $N\to \infty$, with $\zeta$ fixed, quantum fluctuations are not negligible.
Figure~\ref{fig-deltaMcoherent} shows that the relative volume fluctuation of the classical horizon is not negligible for $N\to \infty$ because $\delta V_{\rm C}/V_{\rm C}=\delta \langle M\rangle_N/\langle M\rangle_N$ holds.
In addition, the relative volume fluctuation of the quantum horizon $\delta V_{\rm Q}/V_{\rm Q}$ is not negligible in this limit, either.
This is seen in Fig.~\ref{fig-deltaV(all)zeta=1-10-40}, which shows that the relative volume fluctuation $\delta\langle V_{\rm E}\rangle_N/\langle V_{\rm E}\rangle_N$ is in the range of $\min_{t\in\mathbb{R}}\omega\langle {\bar x}^2\rangle_N\le \omega\langle {\bar x}^2\rangle_N\le \max_{t\in\mathbb{R}}\omega\langle {\bar x}^2\rangle_N$ for $\zeta=1,10,40$.
Since $\omega\langle {\bar x}^2\rangle_N$ is proportional to $\langle V_{\rm E}\rangle_N$ and its dynamics is simple oscillation as shown in Eqs.~(\ref{<X2>-II1})--(\ref{<X2>-II2}), $\delta V_{\rm Q}/V_{\rm Q}$ is given by the value of $\delta\langle V_{\rm E}\rangle_N/\langle V_{\rm E}\rangle_N$ at $\omega\langle {\bar x}^2\rangle_N=\max_{t\in\mathbb{R}}\omega\langle {\bar x}^2\rangle_N$. 
Therefore, Fig.~\ref{fig-deltaV(all)zeta=1-10-40} shows that $\delta V_{\rm Q}/V_{\rm Q}$ does not become small for $N\to\infty$.
Lastly, Table~\ref{tab:V-zeta1} shows the values of $V_{\rm Q}/V_{\rm C}$ for $N=0,1,\cdots, 5$ with $\zeta=1$ and $\zeta=20$.
Table~\ref{tab:V-zeta1} implies that classical and quantum horizons do not coincide for $N\to\infty$.
\begin{table}[htb]
\begin{center}
  \begin{tabular}{|l|c|c|c|c|c|c|} \hline
    \multicolumn{1}{|l|}{} & \multicolumn{2}{c|}{$\zeta=1$} & \multicolumn{2}{c|}{$\zeta=20$}\\ \hline \hline
     &$\Psi=\Psi^{(+)}_{{\rm II}(N)}$ & $\Psi=\Psi^{(-)}_{{\rm II}(N)}$ &$\Psi=\Psi^{(+)}_{{\rm II}(N)}$ & $\Psi=\Psi^{(-)}_{{\rm II}(N)}$  \\ \hline
    $N=0$  & 0.8420 & 0.6580 & 0.9762 & 0.9762  \\ \hline
    $N=1$  & 0.5109 & 0.8222 & 0.9348 & 0.9348   \\ \hline
    $N=2$  & 0.7323 & 0.5069 & 0.9000 & 0.9000   \\ \hline
    $N=3$  & 0.5643 & 0.5604 & 0.8704 & 0.8704   \\ \hline
    $N=4$  & 0.5382 & 0.6574 & 0.8449 & 0.8448  \\ \hline
    $N=5$  & 0.7238 & 0.5736 & 0.8228 & 0.8224 \\ \hline
  \end{tabular}
  \caption{\label{tab:V-zeta1} Values of $V_{\rm Q}/V_{\rm C}$ with $\zeta=1$ and $\zeta=20$ for $N=0,1,\cdots,5$.}
\end{center}
\end{table}

\section{Summary and future prospects}
\label{sec:summary}

In this paper we have studied the quantum dynamics of  toroidal AdS black holes in the framework of throat quantization pioneered by Louko and M\"akel\"a~\cite{LM96}. The resulting Schr\"odinger equation is equivalent to that of  a quantum harmonic oscillator on the half line.
While the classical dynamics represented by the Hamiltonian is equivalent to the dynamics of the wormhole throat in the maximally extended black-hole spacetime, the time $t$ is by construction the proper time at one AdS infinity in the spacetime.
Therefore, all the expectation values of physical/geometrical quantities are for an observer at rest at AdS infinity.

The Hamiltonian operator in the Hilbert space ${\cal L}^2([0,\infty))$ admits a one-parameter family of boundary conditions at the location of the classical singularity $x=0$ with extension parameter $L$.
For any value of $L$, the evolution of the wave function is unitary, and the classical curvature singularity at the center is resolved quantum mechanically and replaced by a big bounce. We were able to construct exact time-dependent wave functions that satisfy either Dirichlet ($L=0$) or Neumann ($L=\infty$) boundary conditions. 
The existence of such exact time-dependent solutions for this class of black-hole spacetimes provides a useful framework for studying conceptual issues surrounding the quantum mechanics of black holes, such as mass uncertainty and the definition of the quantum horizon in the dynamical setting.

We presented two distinct classes of exact time-dependent quantum states.  Both cases clearly exhibit dynamical singularity resolution: The probability amplitudes, $|\Psi|^2$, oscillate between the origin and a maximum ``near" the classical turning point. We evaluated analytically the mass fluctuations and examined the dynamics and fluctuations of two definitions for the horizon of a quantum black hole, the first obtained from the classical expression in terms of the mass, and the second in terms of a suitably defined Euclidean volume. 

The first class, wave function I, is fully dynamical in that the shape of the probability amplitude changes with time throughout the evolution. It is generally spread out far from the origin and becomes more focused near the origin. As a result, the quantum fluctuations are large for most parameter values, and a semi-classical limit is not realized except in the stationary limit. The second class, wave function II, provides a generalization to the half line of coherent states of the harmonic oscillator on the whole line. The probability amplitudes are shape preserving far from the origin and, under the circumstances where the kinetic term energy dominates the potential energy (i.e., $\zeta\ll N$), quantum fluctuations become small, and the classical and quantum horizons coincide so that they are able to represent semi-classical black-hole states. As expected the semi-classical, shape-preserving nature of the solutions breaks down near the origin due to the boundary conditions that provide the mechanism for singularity resolution.

We have shown that the stationary states are fully quantum even in the large-mass limit $N\to\infty$, in the sense that the classical and quantum horizons do not coincide and quantum fluctuations do not become small.
On the other hand, quantum effects are very much suppressed when the states are highly dynamical, i.e., far from the stationary states. This suggests that semi-classical configurations are realized in this limit.
Since our  results were obtained only within two subspaces of the full parameter space of the exact time-dependent wave functions, it is still an open question as to which of the properties we observed are generic.
In addition, it would be interesting to study time-dependent states for asymptotically flat black holes.
We will address these problems elsewhere.

\subsection*{Acknowledgments}
H.~M. thanks Shunichiro Kinoshita for useful comments after the 70th JSPS Annual Meeting at Waseda University.
H.~M. thanks the Department of Physics, University of Winnipeg, and the Winnipeg Institute for Theoretical Physics for their kind hospitality and support while part of this work was carried out. 
G.~K. is grateful to the Natural Sciences and Engineering Research Council for support.

\appendix

\section{Some integrals with Hermite polynomials}
In this appendix, we show useful integral formulae with the Hermite polynomials which are used in the main text.

\subsection{Hermite polynomials}
The definition of the Hermite polynomials $H_N(y)$ characterized by an integer $N=0,1,2,\cdots$ is
\begin{align}
H_N(y)=(-1)^Ne^{y^2}\frac{\D^N}{\D y^N}e^{-y^2},
\end{align}
which satisfies 
\begin{align}
&H_N'=2yH_N-H_{N+1},\label{H-recursion1}\\
&H_N'=2NH_{N-1},\label{H-recursion2}
\end{align}
where a prime denotes differentiation with respect to $y$.
The first nine Hermite polynomials are given by 
\begin{align}
H_0(y)=&1, \nonumber \\
H_1(y)=&2y, \nonumber \\
H_2(y)=&4y^2-2, \nonumber \\
H_3(y)=&8y^3-12y, \nonumber \\
H_4(y)=&16y^4-48y^2+12, \nonumber \\
H_5(y)=&32y^5-160y^3+120y, \nonumber \\
H_6(y)=&64y^6-480y^4+720y^2-120, \nonumber \\
H_7(y)=&128y^7-1344y^5+3360y^3-1680y, \nonumber \\
H_8(y)=&256y^8-3584y^6+13440y^4-13440y^2+1680. \nonumber
\end{align}
The orthogonality condition is 
\begin{align}
\int_{-\infty}^\infty H_M(y)H_N(y)e^{-y^2}\D y=\sqrt{\pi}2^NN!\delta_{MN}. \label{Hermite-orthogonal}
\end{align}

\subsection{Integrals with Hermite polynomials}
We are going to find the form of the following integral: 
\begin{align}
A_{q(N)}:=\int_{0}^\infty y^q H_N(y)^2e^{-y^2}\D y,\label{AqN}
\end{align}
where $q=0,1,2,3,\cdots$.
The value for $N=0$ is easy to obtain:
\begin{align}
A_{q(0)}=\int_{0}^\infty y^q e^{-y^2}\D y=\frac12\Gamma\biggl(\frac{q+1}{2}\biggl),\label{Aq0}
\end{align}
where $\Gamma(x)$ is the gamma function.
In order to obtain the form for general $N$, the following lemma must be useful:
\begin{lm}
\label{lm:recursion}
$A_{q(N)}$ satisfies the following recursion relation:
\begin{align}
A_{q(N)}=2(q+1)A_{q(N-1)}+4(N-1)^2A_{q(N-2)}.\label{AqN-recursion}
\end{align}
\end{lm}
\noindent
{\it Proof}. 
Using the properties (\ref{H-recursion1}) and (\ref{H-recursion2}), we obtain
\begin{align}
A_{q(N)}=&\int_{0}^{\infty} y^qe^{-y^2}(2yH_{N-1}-H_{N-1}')^2\D y \nonumber \\
=&\int_{0}^{\infty} y^qe^{-y^2}\biggl\{4y^2H_{N-1}^2-8(N-1)yH_{N-1}H_{N-2} +4(N-1)^2H_{N-2}^2\biggl\}\D y.\label{AqN-derive1}
\end{align}
Using integration by parts, we rewrite the first term in the last expression as
\begin{align}
4\int_{0}^{\infty} y^{q+2}e^{-y^2}H_{N-1}^2\D y=&-2\int_{0}^{\infty} y^{q+1}(e^{-y^2})'H_{N-1}^2\D y \nonumber \\
=&-2\biggl[y^{q+1}e^{-y^2}H_{N-1}^2\biggl]_0^{\infty} \nonumber \\
&~~+2\int_{0}^{\infty} e^{-y^2}\biggl\{(q+1)y^qH_{N-1}^2 +2y^{q+1}H_{N-1}H_{N-1}'\biggl\}\D y  \nonumber  \\
=&2\int_{0}^{\infty} \biggl\{(q+1)y^qe^{-y^2}H_{N-1}^2 +4(N-1)y^{q+1}e^{-y^2}H_{N-1}H_{N-2}\biggl\}\D y,\label{AqN-derive2}
\end{align}
where we use Eq.~(\ref{H-recursion2}) at the last equality.
Substituting Eq.~(\ref{AqN-derive2}) into Eq.~(\ref{AqN-derive1}), we obtain Eq.~(\ref{AqN-recursion}).
\qed

\bigskip

The form of $A_{q(N)}$ satisfying the recursion relation (\ref{AqN-recursion}) is determined up to a constant factor and this factor is determined by Eq.~(\ref{Aq0}).
Although we are not able to solve the recursion relation (\ref{AqN-recursion}) for general $q$ and $N$, we present the explicit forms of $A_{q(N)}$ for several values of even $q$\footnote{We first obtained these expressions using Wolfram Alpha and then confirmed that they satisfy the recursion relations (\ref{AqN-recursion}) and Eq.~(\ref{Aq0}).}:
\begin{align}
A_{0(N)}=&\frac12\sqrt{\pi}2^NN!,\label{A0N}\\
A_{2(N)}=&\frac{1}{4}\sqrt{\pi}2^NN!(2N+1),\label{A2N}\\
A_{4(N)}=&\frac{3}{8}\sqrt{\pi}2^NN!(2N^2+2N+1),\label{A4N}\\
A_{6(N)}=&\frac{5}{16}\sqrt{\pi}2^NN!(2N+1)(2N^2+2N+3),\\
A_{8(N)}=&\frac{35}{32}\sqrt{\pi}2^NN!\left\{2N(N+1)(N^2+N+4)+3\right\}.
\end{align}

\section{Details of computation: Whole-line case}
\label{app:wholeline}
In this appendix, we present the details of derivations of several expectation values and their uncertainty for the wave function (\ref{six-solution}) on the whole line ${\bar x}\in(-\infty,\infty)$.

\subsection{$\langle {\bar x}\rangle_N$ and $\langle p\rangle_N$ and their uncertainties}
We will often use the integral
\begin{align}
\langle {\bar x}^q\rangle_N=&\mu\beta\times \frac{1}{2^NN!\sqrt{\pi}\mu} \int_{-\infty}^\infty {\bar x}^qe^{-(\beta(t){\bar x}+\varepsilon(t))^2}H_N(\beta(t){\bar x}+\varepsilon(t))^2\D {\bar x} \nonumber \\
=&\frac{1}{2^NN!\sqrt{\pi}\beta^q}\int_{-\infty}^\infty (y-\varepsilon)^qe^{-y^2}H_N(y)^2\D y
\end{align}
for real number $q$, where $y=\beta{\bar x}+\varepsilon$.
Equations~(\ref{A0N})--(\ref{A4N}) show
\begin{align}
\int_{-\infty}^\infty e^{-y^2}H_N(y)^2\D y=&\sqrt{\pi}2^NN!,\label{int-H0} \\
\int_{-\infty}^\infty y^2e^{-y^2}H_N(y)^2\D y=&\frac{1}{2}\sqrt{\pi}2^NN!(2N+1),\label{int-H2}\\
\int_{-\infty}^\infty y^4e^{-y^2}H_N(y)^2\D y=&\frac{3}{4}\sqrt{\pi}2^NN!(2N^2+2N+1).\label{int-H4}
\end{align}
Using them, we obtain
\begin{align}
\langle {\bar x}\rangle_N=&-\frac{\varepsilon}{\beta},\label{exp-x-full}\\
\langle {\bar x}^2\rangle_N=&\frac{(2N+1)+2\varepsilon^2}{2\beta^2},\\
\langle {\bar x}^4\rangle_N=&\frac{3(2N^2+2N+1)+12\varepsilon^2(2N+1)+4\varepsilon^4}{4\beta^4}.
\end{align}
Then, the relative fluctuations are given by 
\begin{align}
\frac{\delta \langle {\bar x}\rangle_N}{\langle {\bar x}\rangle_N}=&\frac{\sqrt{\langle {\bar x}^2\rangle_N-\langle {\bar x}\rangle_N^2}}{\langle {\bar x}\rangle_N} \nonumber \\
=&-\frac{1}{\varepsilon}\sqrt{\frac{2N+1}{2}},\\
\frac{\delta \langle {\bar x}^2\rangle_N}{\langle {\bar x}^2\rangle_N}=&\frac{\sqrt{\langle {\bar x}^4\rangle_N-\langle {\bar x}^2\rangle_N^2}}{\langle {\bar x}^2\rangle_N}\nonumber \\
 =&\frac{\sqrt{2(N^2+N+1)+8(2N+1)\varepsilon^2}}{(2N+1)+2\varepsilon^2}.
\end{align}

In order to compute the expectation value of the momentum ${\hat p}=-i\hbar\partial/\partial{\bar x}$ and its fluctuation, we use the following quantities:
\begin{align}
\frac{\partial \Psi_N}{\partial {\bar x}}=&\biggl\{i\left(2\alpha{\bar x}+\delta\right)-\beta(\beta{\bar x}+\varepsilon) +\frac{2N\beta H_{N-1}(\beta{\bar x}+\varepsilon)}{H_N(\beta{\bar x}+\varepsilon)}\biggl\}\Psi_N,
\end{align}
where we have used Eq.~(\ref{H-recursion2}).
Using this, we obtain
\begin{align}
\langle p\rangle_N=&-\frac{i\hbar }{2^NN!\beta\sqrt{\pi}} \nonumber \\
&\times \int_{-\infty}^{\infty} \biggl[H_N(y)^2\biggl\{i\biggl(2\alpha(y-\varepsilon)+\beta\delta\biggl)-\beta^2y\biggl\}+2N\beta^2 H_{N-1}(y)H_N(y)\biggl]e^{-y^2}\D y \nonumber \\
=&\hbar\biggl(\delta-\frac{2\alpha\varepsilon}{\beta}\biggl),\label{exp-p-full}
\end{align}
where $y=\beta{\bar x}+\varepsilon$.
Now, in order to derive the momentum fluctuation $\delta \langle p\rangle_N$, we compute $\langle p^2\rangle_N$ as
\begin{align}
\langle p^2\rangle_N=&-\hbar^2\mu\beta\int_{-\infty}^{\infty}\Psi_N^*\frac{\partial^2 \Psi_N}{\partial {\bar x}^2}\D{\bar x} \nonumber \\
=&-\hbar^2\mu\beta\biggl\{\biggl[\Psi^*_N\frac{\partial \Psi_N}{\partial {\bar x}}\biggl]_{-\infty}^{\infty}-\int_{-\infty}^{\infty}\frac{\partial \Psi^*_N}{\partial {\bar x}}\frac{\partial \Psi_N}{\partial {\bar x}}\D{\bar x}\biggl\} \nonumber \\
=&\frac{\hbar^2}{2^NN!\sqrt{\pi}}\int_{-\infty}^{\infty}\biggl[\beta^{-2}\biggl\{4\alpha^2 y^2+(\beta\delta-2\alpha \varepsilon)^2\biggl\}H_N(y)^2 \nonumber \\
&~~~~+\beta^2\biggl\{4N^2H_{N-1}(y)^2-4NyH_N(y) H_{N-1}(y) + y^2H_N(y)^2\biggl\}\biggl]e^{-y^2}\D y.
\end{align}
Using the formula
\begin{align}
\int_{-\infty}^\infty 4Ny H_{N-1}(y)H_N(y) e^{-y^2}\D y =&\int_{-\infty}^\infty y (H_{N}^2)' e^{-y^2}\D y \nonumber \\
=&\biggl[y H_{N}^2 e^{-y^2}\biggl]_{-\infty}^\infty-\int_{-\infty}^\infty (y  e^{-y^2})'H_{N}^2\D y \nonumber \\
=&-\int_{-\infty}^\infty( 1-2y^2)H_{N}^2e^{-y^2}\D y
\end{align}
in the last expression, we finally obtain
\begin{align}
\langle p^2\rangle_N=\hbar^2\biggl\{\frac{2N+1}{2\beta^2}(4\alpha^2+\beta^{4})+\biggl(\delta-\frac{2\alpha\varepsilon}{\beta}\biggl)^{2}\biggl\}.
\end{align}
The fluctuations of the position and momentum are given by  
\begin{align}
\delta \langle {\bar x}\rangle_N:=&\sqrt{\langle {\bar x}^2\rangle_N-\langle {\bar x}\rangle_N^2}=\sqrt{\frac{2N+1}{2\beta^2}} \nonumber \\
=&\sqrt{\frac{2N+1}{2{\bar\beta}_0^2}\biggl\{\frac{{\bar\beta}_0^4}{\omega^2}\sin^2\omega t+(2\alpha_0\sin \omega t+\cos \omega t)^2\biggl\}},\\
\delta \langle p\rangle_N:=&\sqrt{\langle p^2\rangle_N-\langle p\rangle_N^2}=\sqrt{\frac{\hbar^2(2N+1)}{2\beta^2}(4\alpha^2+\beta^{4})} \nonumber \\
=&\sqrt{\frac{(2N+1)\hbar^2}{2{\bar\beta}_0^2}\biggl\{{\bar\beta}_0^4\cos^2\omega t+\omega^2(\sin\omega t-2\alpha_0\cos\omega t)^2\biggl\}},
\end{align}
which leads the uncertainty relation:
\begin{align}
&\delta \langle {\bar x}\rangle_N\delta \langle p\rangle_N \nonumber \\
&~~=\frac{(2N+1)\hbar}{4{\bar\beta}_0^2\omega}\biggl\{(4\alpha_0^2\omega^2+{\bar\beta}_0^4+\omega^2)^2-\biggl((4\alpha_0^2\omega^2+{\bar\beta}_0^4-\omega^2)\cos2\omega t-4\alpha_0\omega^2\sin2\omega t\biggl)^2\biggl\}^{1/2}, 
\end{align}
where the expression inside the square root is positive definite.

\subsection{Energy expectation value}
\label{app:energy-full}
For the wave function (\ref{six-solution}) on the whole line ${\bar x}\in(-\infty,\infty)$, the energy expectation value is given by 
\begin{align}
\langle E\rangle_N=&\biggl(N+\frac12\biggl)\hbar\Omega+\frac12\hbar\eta \label{<E>-full-result}\\
=&\biggl(N+\frac12\biggl)\hbar\Omega+\frac{\hbar}{2}\biggl(\omega^2\langle {\bar x}\rangle_N^2+\hbar^{-2}\langle p\rangle_N^2\biggl),
\end{align}
where we have used Eqs.~(\ref{exp-x-full}) and (\ref{exp-p-full}) at the last equality and 
\begin{align}
\Omega:=&\frac{{\bar\beta}_0^4+4\alpha_0^2\omega^2+\omega^2}{2{\bar\beta}_0^2},\label{Omega-app}\\
\eta:=&\frac{(2\alpha_0\varepsilon_0\omega-{\bar\beta}_0{\bar \delta}_0)^2+\varepsilon_0^2\omega^2}{{\bar\beta}_0^{2}}.\label{eta-app}
\end{align}
The energy is equally spaced $\Delta \langle E\rangle_N:=\langle E\rangle_{N+1}-\langle E\rangle_N=\hbar\Omega$.
The energy fluctuation $\delta \langle E\rangle_N:=\sqrt{\langle E^2\rangle_N-\langle E\rangle_N^2}$ is time dependent in general, and its expression is quite complicated; however, it is constant in the shape-preserving state, as shown by Eq.~(\ref{E-uncertain-full-coherent}).

The expression (\ref{<E>-full-result}) is obtained in the following manner.
Using Eqs.~(\ref{t-derivative1})--(\ref{t-derivative7}), we write the time derivative of $\Psi_N$ as
\begin{align}
\frac{\partial \Psi_N}{\partial t}=&\biggl\{-\alpha+\frac12i(\beta^4-4\alpha^2-\omega^2){\bar x}^2 +i(\beta^3\varepsilon-2\alpha\delta){\bar x}+\frac12i(\beta^2\varepsilon^2-\delta^2) \nonumber \\
&-i\frac12(2N+1)\beta^2+(\beta{\bar x}+\varepsilon)(2\beta\alpha{\bar x}+\beta\delta) -\frac{2\beta\alpha{\bar x}+\beta\delta}{H_N(\beta{\bar x}+\varepsilon)}\frac{\D H_N(\beta{\bar x}+\varepsilon)}{\D (\beta{\bar x}+\varepsilon)}\biggl\}\Psi_N. \label{dotPsi}
\end{align}
Then, we compute $\langle E\rangle_N$ as
\begin{align}
\langle E\rangle_N=&\frac{\langle \Psi_N|{\hat H}\Psi_N\rangle}{\langle \Psi_N|\Psi_N\rangle} \nonumber \\
=&\mu\beta\times i\hbar \int_{-\infty}^\infty \Psi_N^*\frac{\partial \Psi_N}{\partial t}\D{\bar x} \nonumber \\
=&\frac{\hbar}{2^NN!\sqrt{\pi}}\int_{-\infty}^\infty \biggl\{-i\alpha -\frac12(\beta^4-4\alpha^2-\omega^2)(y-\varepsilon)^2\beta^{-2} \nonumber \\
& -(\beta^3\varepsilon-2\alpha\delta)(y-\varepsilon)\beta^{-1}-\frac12(\beta^2\varepsilon^2-\delta^2) \nonumber \\
&+\frac12(2N+1)\beta^2+iy\biggl(2\alpha(y-\varepsilon)+\beta\delta\biggl)-i\frac{2\alpha(y-\varepsilon)+\beta\delta}{H_N(y)}\frac{\D H_N(y)}{\D y}\biggl\}e^{-y^2}H_N(y)^2\D y, 
\end{align}
where $y:=\beta{\bar x}+\varepsilon$.
Using integration by parts, we can rewrite the last term in the above expression as
\begin{align}
\int_{-\infty}^\infty \biggl\{2\alpha(y-\varepsilon)+\beta\delta\biggl\}H_N'H_Ne^{-y^2}\D y =&\int_{-\infty}^\infty \frac12\biggl\{2\alpha(y-\varepsilon)+\beta\delta\biggl\}(H_N^2)'e^{-y^2}\D y \nonumber \\
=&\biggl[\frac12\biggl\{2\alpha(y-\varepsilon)+\beta\delta\biggl\} H_N^2e^{-y^2}\biggl]_{-\infty}^\infty \nonumber \\
&-\int_{-\infty}^\infty H_N^2\biggl(\frac12\biggl\{2\alpha(y-\varepsilon)+\beta\delta\biggl\}e^{-y^2}\biggl)'\D y \nonumber \\
=&-\int_{-\infty}^\infty \biggl(\alpha-y\biggl\{2\alpha(y-\varepsilon)+\beta\delta\biggl\}\biggl)H_N^2e^{-y^2}\D y,
\end{align}
where a prime denotes the derivative with respect to $y$.
Finally, by the integral formulae (\ref{int-H0}) and (\ref{int-H2}), we obtain the energy spectrum (\ref{<E>-full-result}).

\subsection{Shape-preserving state}
The shape-preserving state on the whole line is realized for $\alpha_0=0$ and ${\bar\beta}_0=\sqrt{\omega}$, with which we have
\begin{align}
\mu(t)=&\mu_0,\quad \alpha(t)=0,\\
\beta(t)=&\sqrt{\omega},\quad \gamma(t)=\gamma_0-\frac12\omega t,\\
\delta(t)=&{\bar\delta}_0\cos \omega t+\varepsilon_0\sqrt{\omega}\sin \omega t=\sqrt{\omega\zeta}\sin (\omega t+\theta_1),\\
\varepsilon(t)=&\varepsilon_0\cos \omega t-{\bar\delta}_0\sin \omega t/\sqrt{\omega}=\sqrt{\zeta}\sin (\omega t+\theta_2),\\
\kappa(t)=&\kappa_0-\frac{\varepsilon_0{\bar\delta}_0\sin^2 \omega t}{\sqrt{\omega}} +\frac{(\varepsilon_0^2\omega-{\bar\delta}_0^2)\sin 2\omega t}{4\omega },
\end{align}
where $\theta_1:=\arctan({\bar\delta}_0/\varepsilon_0\sqrt{\omega})$, $\theta_2:=\arctan(-\varepsilon_0\sqrt{\omega}/{\bar\delta}_0)$, and 
\begin{align}
\zeta:=\varepsilon_0^2+\frac{{\bar\delta}_0^2}{\omega}.\label{def-zeta-app}
\end{align}
Equations (\ref{Omega-app}) and (\ref{eta-app}) show that $\Omega=\omega$ and $\eta=\omega\zeta$ hold, and hence the energy expectation value reduces to
\begin{align}
\langle E\rangle_N=\biggl(N+\frac12\biggl)\hbar\omega+\frac12\hbar\omega\zeta.\label{<E>-full-coherent}
\end{align}
From Eq.~(\ref{uncertainty-full}), the uncertainty relation in the shape-preserving states is constant:
\begin{align}
\delta \langle {\bar x}\rangle_N\delta \langle p\rangle_N=\hbar\biggl(N+\frac12\biggl). 
\end{align}

Now, let us compute the energy uncertainty in the shape-preserving states.
For this purpose, we need to compute the following quantity:
\begin{align}
\langle E^2\rangle_N:=&\frac{\langle \Psi_N|{\hat H}^2\Psi_N\rangle}{\langle \Psi_N|\Psi_N\rangle} \nonumber \\
=&-\hbar^2\mu\beta\int_{-\infty}^\infty\Psi_N^*\frac{\partial^2\Psi_N}{\partial t^2}\D{\bar x}  \nonumber \\
=&i\hbar\mu\beta\frac{\partial}{\partial t}\int_{-\infty}^\infty\Psi_N^*\biggl(i\hbar\frac{\partial\Psi_N}{\partial t}\biggl)\D{\bar x} +\hbar^2\mu\beta\int_{-\infty}^\infty\frac{\partial\Psi_N^*}{\partial t}\frac{\partial\Psi_N}{\partial t}\D{\bar x} \nonumber \\
=&i\hbar\mu\beta\frac{\partial}{\partial t}\biggl(\mu_0{\bar\beta}_0\langle E\rangle_N\biggl)+\hbar^2\mu\beta\int_{-\infty}^\infty\frac{\partial\Psi_N^*}{\partial t}\frac{\partial\Psi_N}{\partial t}\D{\bar x} \nonumber \\
=&\hbar^2\mu\beta\int_{-\infty}^\infty\frac{\partial\Psi_N^*}{\partial t}\frac{\partial\Psi_N}{\partial t}\D{\bar x}, \label{E^2-full-1}
\end{align}
where we use Eq.~(\ref{<E>-full-result}).
By Eq.~(\ref{dotPsi}), the integral in the last expression becomes
\begin{align}
\int_{-\infty}^\infty \frac{\partial \Psi^*_N}{\partial t}\frac{\partial \Psi_N}{\partial t}\D{\bar x}=&\frac{\beta\delta^2}{2^NN!\mu \sqrt{\pi}}\int_{-\infty}^\infty \biggl\{\biggl((\beta^2\delta^{-2}\varepsilon^2+1) y^2 \nonumber \\
&+\frac{\{\beta^2\varepsilon^2+\delta^2+(2N+1)\beta^2\}^2}{4\beta^{2}\delta^{2}}\biggl)H_N^2 -y(H_N^2)'+4N^2H_{N-1}^2\biggl\}e^{-y^2}\D y,
\end{align}
where $y:=\beta{\bar x}+\varepsilon$ and a prime denotes the derivative with respect to $y$.
Using the following expression for the last term,
\begin{align}
-\int_{-\infty}^\infty y(H_N^2)'e^{-y^2}\D y=& -\biggl[ yH_N(y)^2e^{-y^2}\biggl]_{-\infty}^\infty+\int_{-\infty}^\infty (ye^{-y^2})'H_N^2\D y \nonumber \\
=&\int_{-\infty}^\infty (1-2y^2)e^{-y^2}H_N^2\D y
\end{align}
and Eqs.~(\ref{int-H0}) and (\ref{int-H2}), we obtain
\begin{align}
\int_{-\infty}^\infty \frac{\partial \Psi^*_N}{\partial t}\frac{\partial \Psi_N}{\partial t}\D{\bar x}=&\frac{\beta}{\mu}\biggl\{\frac12(2N+1)(\beta^2\varepsilon^2+\delta^2) +\frac{\{\beta^2\varepsilon^2+\delta^2+(2N+1)\beta^2\}^2}{4\beta^{2}}\biggl\}.
\end{align}
Hence, $\langle E^2\rangle_N$ is written in terms of $\zeta$ as
\begin{align}
\langle E^2\rangle_N=&\frac14\hbar^2\omega^2\biggl\{2(2N+1)\zeta+(\zeta+2N+1)^2\biggl\} .
\end{align}
Using this expression and Eq.~(\ref{<E>-full-coherent}), we finally obtain the energy uncertainty for the shape-preserving states:
\begin{align}
\delta \langle E\rangle_N:=&\sqrt{\langle E^2\rangle_N-\langle E\rangle_N^2} =\hbar\omega\sqrt{\biggl(N+\frac12\biggl)\zeta}. \label{E-uncertain-full-coherent}
\end{align}

\section{Details of computation: Half-line case}
\label{app:halfline}
In this appendix, we present the details of derivations of several expectation values and their uncertainty for the wave function I, given by Eq.~(\ref{sol-first}), on the half line ${\bar x}\in[0,\infty)$.

\subsection{Mass expectation value and its uncertainty}
\label{app:mass}

Here we will show that the mass expectation value is given by 
\begin{align}
\langle M\rangle_N:=\frac{\langle \Psi_{{\rm I}(N)}|{\hat H}\Psi_{{\rm I}(N)}\rangle}{\langle \Psi_{{\rm I}(N)}|\Psi_{{\rm I}(N)}\rangle} =\biggl(N+\frac12\biggl)\hbar\Omega, \label{mass-exp}
\end{align}
where the constant $\Omega$ is defined by Eq.~(\ref{Omega-app}).
The derivation is similar to the case of $\langle E\rangle_N$ shown in Appendix~\ref{app:energy-full} but with $\varepsilon_0={\bar\delta}_0=0$ (and hence $\varepsilon(t)=\delta(t)=0$).
For derivation, we use Eq.~(\ref{s-norm-half}) for the denominator in (\ref{mass-exp}) and the integral formulae (\ref{A0N}) and (\ref{A2N}) on the half line.

Next, let us show the following mass uncertainty:
\begin{align}
\delta \langle M\rangle_N:=&\sqrt{\langle M^2\rangle_N-\langle M\rangle_N^2} =\sqrt{\frac{(N^2+N+1)\hbar^2(\Omega^2-\omega^2)}{2}}. \label{mass-exp-square}
\end{align}
In order to derive this result, we need to compute $\langle M^2\rangle_N(t)$.
The derivation is similar to the case of $\langle E^2\rangle_N$ on the whole line shown in Appendix~\ref{app:energy-full}.

Similarly to Eq.~(\ref{E^2-full-1}), we obtain
\begin{align}
\langle M^2\rangle_N=&\frac{\langle \Psi_{{\rm I}(N)}|{\hat H}^2\Psi_{{\rm I}(N)}\rangle}{\langle \Psi_{{\rm I}(N)}|\Psi_{{\rm I}(N)}\rangle} =2\hbar^2\mu\beta \int_0^\infty\frac{\partial\Psi_{{\rm I}(N)}^*}{\partial t}\frac{\partial\Psi_{{\rm I}(N)}}{\partial t}\D{\bar x},
\end{align}
where we use Eq.~(\ref{mass-exp}) together with Eq.~(\ref{s-norm-half}).
Using Eq.~(\ref{dotPsi}) with $\varepsilon(t)=\delta(t)=0$, we obtain
\begin{align}
\frac{\partial \Psi_{{\rm I}(N)}^*}{\partial t}\frac{\partial \Psi_{{\rm I}(N)}}{\partial t}=&\frac{e^{-\beta^2{\bar x}^2}}{2^NN!\mu\sqrt{\pi}}\biggl\{\alpha^2\biggl((1-2\beta^2{\bar x}^2)H_N(\beta{\bar x})+4N{\bar x}\beta H_{N-1}(\beta{\bar x})\biggl)^2  \nonumber \\
&+\frac14H_N(\beta{\bar x})^2\biggl((\beta^4-4\alpha^2-\omega^2){\bar x}^2-(2N+1)\beta^2\biggl)^2\biggl\}.
\end{align}
By this expression, we have
\begin{align}
\langle M^2\rangle_N=&\frac{\hbar^2}{2^{N-1}N!\sqrt{\pi}}\int_0^\infty e^{-y^2}\biggl\{\alpha^2\biggl((1-2y^2)^2H_N^2+8Ny (1-2y^2)H_{N-1}H_N+16N^2y^2 H_{N-1}^2\biggl)  \nonumber \\
&+\frac14H_N^2\biggl(\beta^{-2}(\beta^4-4\alpha^2-\omega^2)y^2-(2N+1)\beta^2\biggl)^2\biggl\}\D y,\label{M2-exp-step2}
\end{align}
where $y:=\beta{\bar x}$.
By Eq~(\ref{H-recursion2}) and integration by parts, the second term in the above expression becomes
\begin{align}
\int_0^\infty 4Ny(1-2y^2) H_{N-1}H_Ne^{-y^2}\D y =&\int_0^\infty y(1-2y^2)  (H_{N}^2)' e^{-y^2}\D y \nonumber \\
=&\biggl[y (1-2y^2) H_{N}^2 e^{-y^2}\biggl]_0^\infty -\int_0^\infty \left(y(1-2y^2)   e^{-y^2}\right)'H_{N}^2\D y \nonumber \\
=&-\int_0^\infty( 1-8y^2+4y^4)H_{N}^2e^{-y^2}\D y\label{M2-exp-step2}
\end{align}
and hence
\begin{align}
\langle M^2\rangle_N=&\frac{\hbar^2}{2^{N-1}N!\sqrt{\pi}}\int_0^\infty e^{-y^2}\biggl\{16N^2\alpha^2y^2 H_{N-1}^2+y^4H_N^2\biggl(\frac14\beta^{-4}(\beta^4-4\alpha^2-\omega^2)^2-4\alpha^2\biggl)  \nonumber \\
&-y^2H_N^2\biggl(\frac12(2N+1)(\beta^4-4\alpha^2-\omega^2)-12\alpha^2\biggl) +H_N^2\biggl(\frac14(2N+1)^2\beta^4-\alpha^2\biggl)\biggl\}\D y.
\end{align}
Finally, using the integral formulae (\ref{A0N}), (\ref{A2N}), and (\ref{A4N}) and Eq.~(\ref{mass-exp}), we obtain
\begin{align}
\langle M^2\rangle_N=&\frac{3(2N^2+2N+1)}{(2N+1)^2}\langle M\rangle_N^2-\frac{N^2+N+1}{2}\hbar^2\omega^2. \label{M^2-result}
\end{align}
This expression and Eq.~(\ref{mass-exp}) give the mass uncertainty (\ref{mass-exp-square}).

\subsection{Area and volume expectation values and their uncertainties}
\label{app:area}

Here we present the expressions of the expectation values of the surface area $A$ and the Euclidean volume $V_{\rm E}$ and their uncertainties.

For this purpose, we derive the expression of $\langle x^p\rangle_N$ and its uncertainty with a real number $p$.
In terms of the coordinate ${\bar x}$, it is written as
\begin{align}
\langle x^p\rangle_N =&\biggl(\frac{(n-1)^2\hbar\kappa_n^2}{4(n-2)V_{n-2}^{(0)}}\biggl)^{p/2}\langle {\bar x}^p\rangle_N \nonumber \\
=&\biggl(\frac{(n-1)^2\hbar\kappa_n^2}{4(n-2)V_{n-2}^{(0)}}\biggl)^{p/2}\frac{\int_{0}^\infty {\bar x}^p|\Psi_{{\rm I}(N)}({\bar x},t)|^2\D {\bar x}}{\int_{0}^\infty |\Psi_{{\rm I}(N)}({\bar x},t)|^2\D {\bar x}} \nonumber \\
=&\biggl(\frac{(n-1)^2\hbar\kappa_n^2}{4(n-2)V_{n-2}^{(0)}}\biggl)^{p/2}\frac{1}{2^{N-1}N!\sqrt{\pi}\beta^p} \int_{0}^{\infty} y^p e^{-y^2}H_N(y)^2 \D y,\label{<x^p>}
\end{align}
where $y:=\beta{\bar x}$ and we have used Eq.~(\ref{s-norm-half}).
Using the expression
\begin{align}
\langle x^{2p}\rangle_N =&\biggl(\frac{(n-1)^2\hbar\kappa_n^2}{4(n-2)V_{n-2}^{(0)}}\biggl)^{p} \frac{1}{2^{N-1}N!\sqrt{\pi}\beta^{2p}}\int_{0}^{\infty} y^{2p} e^{-y^2}H_N(y)^2 \D y,
\end{align}
we obtain the uncertainty as
\begin{align}
\delta \langle x^p\rangle_N:=&\sqrt{\langle x^{2p}\rangle_N-\langle x^p\rangle_N^2} \nonumber \\
=&\biggl(\frac{(n-1)^2\hbar\kappa_n^2}{4(n-2)V_{n-2}^{(0)}}\biggl)^{p/2}\frac{1}{\beta^p}\sqrt{\frac{\int_{0}^{\infty} y^{2p} e^{-y^2}H_N^2 \D y}{2^{N-1}N!\sqrt{\pi}}-\biggl(\frac{\int_{0}^{\infty} y^{p} e^{-y^2}H_N^2 \D y}{2^{N-1}N!\sqrt{\pi}}\biggl)^2}.\label{delta<x^p>}
\end{align}

The surface area with an areal radius $r(=a)$ is given by $A:=V_{n-2}^{(0)}a^{n-2}=V_{n-2}^{(0)}x^{2(n-2)/(n-1)}$, where $x:=a^{(n-1)/2}$.
Hence, its expectation value and uncertainty are given by
\begin{align}
\langle A\rangle_N=&\biggl(\frac{(n-1)^2\hbar\kappa_n^2}{4(n-2)V_{n-2}^{(0)}}\biggl)^{(n-2)/(n-1)}\frac{V_{n-2}^{(0)}\int_{0}^{\infty} y^{2(n-2)/(n-1)} e^{-y^2}H_N^2 \D y}{2^{N-1}N!\sqrt{\pi}\beta^{2(n-2)/(n-1)}}, \label{area-exp}\\
\delta \langle A\rangle_N =&\biggl(\frac{(n-1)^2\hbar\kappa_n^2}{4(n-2)V_{n-2}^{(0)}}\biggl)^{(n-2)/(n-1)}\frac{V_{n-2}^{(0)}}{\beta^{2(n-2)/(n-1)}} \nonumber \\
&\times\sqrt{\frac{\int_{0}^{\infty} y^{4(n-2)/(n-1)} e^{-y^2}H_N^2 \D y}{2^{N-1}N!\sqrt{\pi}}-\biggl(\frac{\int_{0}^{\infty} y^{2(n-2)/(n-1)} e^{-y^2}H_N^2 \D y}{2^{N-1}N!\sqrt{\pi}}\biggl)^2}. \label{delta-area-exp}
\end{align}
The above expressions show 
\begin{align}
\frac{\delta \langle A\rangle_N}{\langle A\rangle_N}=\sqrt{\frac{2^{N-1}N!\sqrt{\pi}\int_{0}^{\infty} y^{4(n-2)/(n-1)} e^{-y^2}H_N^2 \D y}{(\int_{0}^{\infty} y^{2(n-2)/(n-1)} e^{-y^2}H_N^2 \D y)^2}-1},
\end{align}
which is constant and independent of the parameters.

The Euclidean volume with the areal radius $r(=a)$ is given by 
\begin{align}
V_{\rm E}:=\frac{V_{n-2}^{(0)}}{n-1}a^{n-1}=\frac{V_{n-2}^{(0)}}{n-1}x^2.
\end{align}
Using the integral formulae (\ref{A2N}) and (\ref{A4N}), we obtain the volume expectation value and its uncertainty as
\begin{align}
\langle V_{\rm E}\rangle_N=&\frac{(n-1)\hbar\kappa_n^2}{4(n-2)\beta(t)^2}\biggl(N+\frac12\biggl),\\
\delta \langle V_{\rm E}\rangle_N=&\frac{(n-1)\hbar\kappa_n^2}{4(n-2)\beta(t)^2}\sqrt{\frac{N^2+N+1}{2}}.
\end{align}
The above expressions show
\begin{align}
\frac{\delta \langle V_{\rm E}\rangle_N}{\langle V_{\rm E}\rangle_N}=\sqrt{\frac{2(N^2+N+1)}{(2N+1)^2}}.
\end{align}


\end{document}